\documentclass[twocolumn,showpacs,preprintnumbers,amsmath,amssymb,floatfix]{revtex4}

\usepackage{graphicx}
\usepackage{dcolumn}
\usepackage{bm}
\usepackage{pst-all}      
\usepackage{fancybox}

\newcommand{\bfq}{{\bf q}}
\newcommand{\bfk}{{\bf k}}
\newcommand{\bfr}{{\bf r}}
\newcommand{\bfQ}{{\bf Q}}

\newcounter{Alist}

\newcommand{\ocite}[1]{\cite{#1}}

\newcommand{\ispone}{}
\newcommand{\isptwo}{}

\def\qsgw{QS{\em GW}}

\def\GLDA{{G^{\rm LDA}}}
\def\WLDA{{W^{\rm LDA}}}
\def\ekn{{\varepsilon_{{\bf k}n}}}

\def\ei{\varepsilon_i}

\def\ej{\varepsilon_j}
\def\Ekn{{E_{{\bf k}n}}}
\def\Psikn{\Psi_{{\bf k}n}}
\def\Psiqn{{\Psi_{{\bf q}n}}}
\def\Psiqm{{\Psi_{{\bf q}m}}}
\def\DVo{{\it \Delta}V(\omega)}

\def\HLDA{H^{\rm LDA}}
\def\H0{H^0}

\def\veff{V^{\rm eff}}
\def\vxc{V^{\rm xc}}

\def\vext{V^{\rm ext}}
\def\hVext{\hat{V}^{\rm ext}}
\def\hVeff{\hat{V}^{\rm eff}}

\def\vh{V^{\rm H}}
\def\vgw{V^{GW}}

\def\gwa{$GW$\!A}
\def\hVee{\hat{V}^{\rm ee}}
\def\hVext{\hat{V}^{\rm ext}}
\def\hHk{\hat{H}^{\rm k}}

\newcommand{\req}[1]{\mbox{Eq.~(\ref{#1})}}

\def\scgw{{QS{\em GW}}}

\def\ekn{{\varepsilon_{{\bf k}n}}}
\def\ekm{{\varepsilon_{{\bf k}m}}}

\def\Ekn{{E_{{\bf k}n}}}
\def\Psikn{{\Psi_{{\bf k}n}}}

\def\Psikmstar{{ \Psi_{{\bf k}m}^*} }
\def\Psiknp{{\Psi_{{\bf k}n'}}}

\def\brl{{\bf R}L}
\def\brlp{{{\bf R}'L'}}

\def\tiln{{\tilde{n}}}
\def\tilm{{\tilde{m}}}
\newcommand{\val}{{\rm{VAL}}}
\newcommand{\core}{{\rm{CORE}}}

\newcommand{\CORE}{{CORE}}

\newcommand{\VAL}{{\rm{VAL}}}
\newcommand{\EF}{E_{\rm F}}
\newcommand{\oneshotgw}{1shot-$GW$}

\def\eak{\varepsilon_{\rm a}(\bfk)}
\def\ebk{\varepsilon_{\rm b}(\bfk)}
\def\iDelta{{\it \Delta}}
\def\efermi{\mbox{$E_{\rm F}$}}

\def\connect#1{\leavevmode{\setbox1=\hbox{#1}\copy1%
\raise .2\ht1 \vbox{\moveleft \wd1\vbox{\hrule width \wd1 height .5pt depth 0pt}}%
}}
\def\we{\mbox{$\omega_\varepsilon$}}

\def\GG{$\Gamma$}
\def\x{\mbox{$\times$}}

\def\xccutone{ {\rm xccut1} }
\def\xccuttwo{ {\rm xccut2} }

\def\ftn[#1]{\rlap{\footnotemark[#1]}}

\def\tr{{\rm Tr}}

\def\bQP{{\it bare QP}}
\def\bQPs{{\it bare QPs}}
\def\dQP{{\it dressed QP}}
\def\dQPs{{\it dressed QPs}}
\def\Re{{\rm Re}}

\begin{document}


%

\newpage
\title{Quasiparticle self-consistent $GW$ method; a basis for the independent-particle approximation}
\author{Takao Kotani}
\affiliation{School of Materials, Arizona State University, Tempe, AZ, 85284}
\author{Mark van Schilfgaarde}
\affiliation{School of Materials, Arizona State University, Tempe, AZ, 85284}
\author{Sergey V. Faleev}
\affiliation{Sandia National Laboratories, Livermore, CA 94551}

\date{\today}

\begin{abstract}
We have developed a new type of self-consistent scheme 
within the $GW$ approximation,
which we call quasiparticle self-consistent $GW$ (QS$GW$).  We have shown
that QS$GW$describes energy bands for a wide-range of
materials  rather well, including many where the local-density approximation fails.
QS$GW$ contains physical effects found in other theories such as LDA$+U$,
SIC and $GW$ in a satisfactory manner without many of their drawbacks
(partitioning of itinerant and localized electrons, adjustable parameters,
ambiguities in double-counting, etc.).  We present some theoretical
discussion concerning the formulation of QS$GW$, including a prescription
for calculating the total energy.  We also address several key
methodological points needed for implementation.  We then show convergence
checks and some representative results in a variety of materials.
\end{abstract}


\maketitle
In the 1980's, algorithmic developments and faster computers made it possible
to apply Hedin's $GW$ approximation 
(\gwa) \cite{hedin65} to real materials~\cite{Strinati80,pickett84}.
Especially, Hybertsen and Louie \cite{hybertsen86} first implemented
the \gwa\ within an \emph{ab-initio} framework in a satisfactory manner.
Theirs was a perturbation treatment starting from the Kohn-Sham
eigenfunctions and eigenvalues given in the local density approximation
(LDA) to density functional theory (DFT)\cite{hohenberg64,kohn65}.  We will
denote this approach here as \oneshotgw.  Until now \oneshotgw\ has been
applied to variety of materials, usually in conjunction with the
pseudopotential (PP) approximation.  Quasiparticle (QP) energies so
obtained are in significantly better agreement with experiments than the
LDA Kohn-Sham eigenvalues\cite{ferdi98}.


%
%




However, we have recently shown that \oneshotgw\ has many significant
failings.  Even in simple semiconductors it systematically underestimates
optical gaps\cite{kotani02,Usuda04,Fleszar05,mark06adeq}.  
In general, the quality of results are closely tied to the quality of the LDA starting
point.  For more complicated cases where the LDA eigenfunctions are poor,
\oneshotgw\ can fail even qualitatively\cite{mark06adeq}.

A possible way to overcome this difficulty is to determine
the starting point self-consistently.
The effects of the eigenvalue-only self-consistency 
(keeping the eigenfunctions as given in LDA),
was discussed by Surh, Louie, and Cohen\cite{surh91}. 
Recently, Luo, Ismail-Beigi, Cohen, and Louie \cite{luo02} applied it to 
ZnS and ZnSe, where they showed that 
the band gaps of \oneshotgw\ 3.19~eV and 2.32~eV for ZnS and ZnSe are
increased to 3.64~eV and 2.41~eV by the eigenvalue-only self-consistency 
(see Table~\ref{tab:gap3} also). The differences suggest 
the importance of this self-consistency.
Furthermore, for ZnSe, the value 2.41~eV changes to
2.69~eV when they use eigenfunctions 
given by generalized gradient approximation (GGA).  This difference 
suggests that we may need to look for a means
to determine optimum eigenfunctions for $GW$A.
Aryasetiawan and Gunnarsson applied another kind of self-consistent scheme
to NiO \cite{aryasetiawan95nio}.  They introduced a parameter for the non-local potential 
which affects the unoccupied $e_g$ level, and made it self-consistent.
They showed that the band gap of \oneshotgw\ is $\sim$1~eV, and that it is 
improved to $\sim$5.5~eV by the self-consistency. 

Based on these self-consistency ideas, we have developed 
a new \emph{ab-initio} approach to $GW$
\cite{faleev04,mark06qsgw,chantis06a,chantis06fshell}, which we now call
``quasiparticle self-consistent $GW$'' (\qsgw) method.  
\qsgw\ is a first-principles method that stays within the framework of Hedin's \gwa,
that is, \qsgw\ is a perturbation theory built around some noninteracting
Hamiltonian.  It does not depend on the LDA anymore but rather determines the
optimum noninteracting Hamiltonian in a self-consistent manner.
We have shown that \qsgw\ satisfactorily describes QP energies 
for a wide range of materials. 
Bruneval, Vast and Reining \cite{bruneval06qsgw} implemented 
it in the pseudopotential scheme, and gave some kinds of analysis
including the comparison with the Hartree-Fock method and 
with the Coulomb-hole and Screened exchange (COHSEX) methods.

The present paper begins with a derivation of the fundamental equation of
\qsgw, and some theoretical discussion concerning it
(Sec.~\ref{sec:theory}).  The fundamental equation is derived
from the idea of a self-consistent perturbation.  We also present a
means for computing the total energy through the adiabatic connection
formalism. 
Next, we detail a
number of key methodological points (Sec.~\ref{sec:method}).  The present
implementation is unique in that it makes no pseudopotential or shape
approximation to the potential, and it uses a mixed basis for the response
function, Coulomb interaction, and self-energy, which enables us to
properly treat core states.  The $GW$A methodology is presented along with some
additional points particular to self-consistency.  In
Sec.~\ref{sec:result}, we show some convergence checks, using GaAs as a
representative system.  Then we show how \qsgw\ works by comparing it to
other kinds of \gwa\ for compounds representative of different materials
classes: semiconductors C, Si, SiC, GaAs, ZnS, and ZnSe; oxide
semiconductors ZnO and Cu$_2$O; transition metal monoxides MnO and NiO;
transition metals Fe and Ni.


\section{Theory}
\label{sec:theory}
\subsection{\gwa}
Let us summarize the \gwa~\cite{hedin65,hybertsen86} for later discussion. 
Here we omit spin index for simplicity. 
Generally speaking, we can perform $GW$A from 
some given one-body Hamiltonian $\H0$ written as
\begin{equation}
\H0 = \frac{-\nabla^2}{2m} + \veff(\bfr,\bfr').
\label{eq:defh0}
\end{equation}
The one-particle effective potential $\veff(\bfr,\bfr')$ 
can be non-local, though it is local, i.e.
$\veff(\bfr,\bfr')=\veff(\bfr)\delta(\bfr-\bfr')$
when generated by the usual Kohn-Sham construction.
$\H0$ determines the set of eigenvalues $\{\ei\}$  and eigenfunctions $\{\Psi_i(\bfr)\}$.
From them
we can construct the non-interacting Green's function $G^0$ as
\begin{equation}
G^0\left( \bfr,\bfr',\omega \right) 
= \sum\limits_i {
\frac{\Psi_i(\bfr) \Psi_i^*(\bfr')}
     {{\omega -\ei \pm i \delta }}},
\label{eq:defg}
\end{equation}
where $-i \delta$ is for occupied states, and
$+i\delta$ for unoccupied states.   
Within the RPA (random-phase approximation), the screened Coulomb interaction is
\begin{equation}
 W = \epsilon^{-1} v = \left(1-v \Pi\right)^{-1} v
\label{eq:defw}
\end{equation}
where $\Pi=-iG^0\times{}G^0$ is the proper polarization function,
and $v(\bfr, \bfr')=\frac{e^2}{|\bfr-\bfr'|}$ 
is the bare Coulomb interaction. 
$\epsilon$ denotes the dielectric function.
As seen in e.g., works by Alouani and co-workers \cite{alouani96,arnaud01},
$W$ calculated from a reasonable $\H0$ should 
be in good agreement with experiments, even if $W$ 
does not satisfy the $f$-sum rule because $\H0$ is non-local \cite{alouani96}
(because of the so-called scissors operator).

Hedin's \gwa\ gives the self-energy $\Sigma(\bfr, \bfr'\!\!, \omega)$ as
\begin{equation} 
\label{self_energy}
\Sigma(\bfr, \bfr'\!\!, \omega)
\!=\!\frac{i}{2\pi}\!\!\int\!\!d\omega'
G^0(\bfr,\bfr'\!\!,\omega-\omega')W(\bfr,\bfr'\!\!,\omega')e^{-i\delta\omega'}\!\!.
\end{equation}
From this self-energy,
the external potential $\vext$ from the nuclei,
and the Hartree potential $\vh$
which is calculated from the electron density through $G^0$, 
we obtain an $\omega$-dependent one-body effective potential $\vgw(\omega)$ :
\begin{equation} 
\vgw(\omega) =\vext + \vh + \Sigma(\omega).
\label{vgw}
\end{equation}
Note that $\vh$ is detemined from the density which is calculated
for the non-interacting system specified by $\H0$. 
For simplicity we omit arguments $(\bfr,\bfr')$.
Then the one-body Green function 
is given as 
$G = 1/({-\nabla^2}{2m} + \vgw(\omega))$.
$\vext$ and $\vh$ are local and $\omega$-independent potentials.
Thus the \gwa\ maps $\veff$ to $\vgw(\omega)$.
In other words, the \gwa\ generates a perturbative correction $\DVo$
to the one-particle potential $\veff$, written as
\begin{eqnarray}
\DVo = \vgw(\omega) - \veff .
\label{eq:dvo}
\end{eqnarray}
$\vgw(\omega)$ and $\DVo$ can be regarded as functionals 
of $\veff$ (or $\H0$). 

In the standard \oneshotgw\ with $H^0$ generated by the LDA, $\veff$
is the LDA Kohn-Sham Hamiltonian.  Neglecting off-diagonal terms, the QP
energy (QPE) is
\begin{widetext}
\begin{equation}
\Ekn=\ekn + Z_{\bfk n}[
  \langle\Psikn | \Sigma(\bfr,\bfr',\ekn)   |\Psikn \rangle
- \langle\Psikn |V_{\rm xc}^{\rm{LDA}}(\bfr)|\Psikn \rangle],
\label{eq:e1shot}
\end{equation}
where $Z_{{\bfk}n}$ is the QP renormalization factor:
\begin{equation}
Z_{{\bf k}n}=\left[ 1-\langle\Psikn|
\frac{\partial}{\partial\omega} \Sigma({\bf r},{\bf r}^{\prime},\ekn)
|\Psikn \rangle \right]^{-1} .
\label{eq:defzfac}
\end{equation}
\end{widetext}
Subscripts label the wave vector $\bfk$
and band index $n$.  We will write them later as a compound index, $i \equiv (\bfk, n)$.
Eq.~(\ref{eq:e1shot}) is the customary way QPEs are calculated in $GW$.
However, as we discussed in Ref.~\ocite{mark06adeq}, using $Z$=1 instead of
Eq.~(\ref{eq:defzfac}) is usually a better approximation; see also
Sec.~\ref{sec:result}. 
Chapter 7 of Ref.\ocite{mahan90} presents another
analysis  where $Z$=1 is shown to be a better approximation, in the context of the Fr\"olich Hamiltonian.
In any case, we have to calculate
matrix elements $\langle\Psiknp|\Sigma(\bfr,\bfr',\omega)|\Psikn \rangle$
as accurately and as efficiently as possible
(off-diagonal elements are necessary in the \qsgw\ case, as explained below).

As we showed in Ref.~\ocite{mark06adeq}, 
$\H0$ generated by LDA is not necessarily a good approximation.
[Even the $\H0$ for ``true Kohn-Sham'' Hamiltonian in DFT 
can be a poor descriptor of QP excitation energies
\cite{kotani98}.]  For example, time-reversal
symmetry is automatically enforced because $\veff$ is local (and thus
real).  This symmetry is strongly violated in open $f$-shell systems
\cite{chantis06fshell}.  The bandgap of a relatively simple III-V
semiconductor, InN, is close to zero~\cite{kotani02,Usuda04}; 
also the QP spectrum of NiO is little improved over
LDA~\cite{faleev04}.  A variety of other examples could be cited where
$GW$A starting from $\H0=H^{\rm{LDA}}$ is a poor approximation.
(In contrast, see Sec.~\ref{sec:result} and Ref.~\cite{mark06qsgw} 
to see how 
\qsgw\ gives 
consistently
good agreement with experiment.) 

\subsection{Quasiparticle self-consistent $GW$}
\label{sec:qsgw}
\qsgw\ is a formalism which determines $\veff$ (or $\H0$) 
self-consistently within the $GW$A, without depending on LDA or DFT.
If we have a mapping procedure $\vgw(\omega) \rightarrow
\veff$, we can close the equation to determine $\veff$, i.e. determine
$\veff$ self-consistently by $\veff \rightarrow \vgw(\omega) \rightarrow
\veff \rightarrow \dots$. 
The main idea to determine the mapping is
grounded 
in the concept of the QP. Roughly speaking,
$\veff$ is determined so as to reproduce the QP generated from $\vgw(\omega)$.
In the following, we explain how to determine this $\vgw(\omega)
\rightarrow \veff$, and derive the fundamental \qsgw\ equation
\cite{faleev04,mark06qsgw,chantis06a,chantis06fshell}.

Based on Landau's QP picture, there are fundamental 
one-particle-like excitations denoted as quasiparticles (QP), at least around the Fermi energy \efermi.
The QPEs and QP eigenfunctions (QPeigs), $\{E_i, \Phi_i(\bfr)\}$, are given as 
\cite{hedin65}
\begin{eqnarray}
\left[\frac{-\nabla^2}{2m} + \vext +\vh + {\rm Re}[\Sigma(E_i)] - E_i\right] |\Phi_i\rangle = 0.
\label{eq:qp}
\end{eqnarray}
We refer to the states characterized by these $E_i$ and $\Phi_i(\bfr)$
as the \dQP. Here ${\rm Re}[X]$ means
just take the hermitian part of $X$ so $E_i$ is real
for $E_i$.  
This is irrelevant 
around \efermi\ because the anti-hermitian part of $\Sigma(E_i)$
goes to zero as $E_i \to$ \efermi.
On the other hand, we have another one-particle picture
described by $\H0$; we name these QPs as \bQP{s},
and refer to the QPEs and eigenfunctions corresponding to $\H0$ as $\{\ei, \Psi_i(\bfr)\}$.

Let us consider the difference and the relation of these two kinds of QP.
The \bQP\ is essentially consistent with the Landau-Silin QP picture,  
discussed by, e.g., Pines and Nozieres in Sec. 3.3, Ref.~\ocite{pines66}.  
The \bQP\ interact with each other via the bare Coulomb interaction.
The \bQP\ given by ${\H0}$ evolve into the \dQP\
when the interaction $\hat{H} -\hat{\H0}$ is 
turned on adiabatically.
Here $\hat{H}$ is the total Hamiltonian (See \req{hami}); and the hat 
 signifies that $\hat{H}$ is written in second quantized form. 
 $\hat{\H0}$ and $\H0$ are equivalent.
The \dQP\ consists of the central \bQP\ and an
induced polarization cloud consisting of other \bQP\ ; this view
is compatible with the way interactions are treated in the \gwa.

$\H0$ generating the \bQPs\ represents a virtual reference system 
just for  theoretical convenience.
There is an ambiguity in how to determine $\H0$;
in principle, any $\H0$ can be used if $\hat{H} -\hat{\H0}$ 
could be completely included. However, as we evaluate the difference 
$\hat{H} - \hat{\H0}$ in some perturbation method like $GW$A, 
we must utilize some optimum (or best) $\hat{\H0}$:
$\hat{\H0}$ should be chosen so that the perturbative 
contribution is as small as possible.
A key point remains in how to define a
measure of the size of the perturbation.
We can classify our \qsgw\ method as a self-consistent perturbation method
which self-consistently determines the optinum division 
of $\hat{H}$ into the main part $\H0$ and the residual part $\hat{H} -\hat{\H0}$.
There are various possible choices for the measure; however, here we
take a simple way, by requiring that the two kinds of QPs discussed
in the previous paragraphs correspond as closely as possible.
We choose $\H0$ so as to repoduce the \dQPs.
In other words, we assign the difference of the QPeig (and also the QPE)
between the \bQP\ and the \dQP\ as the measure,
and then we minimize it.
From the physical point of view, this means 
that the motion of the central electron of the \dQP\
is not changed by $\hat{H} - \hat{\H0}$.
Note that $\hat{H} -\hat{\H0}$ contains two kinds of contributions:
not only the Coulomb interaction but also the one-body term $\vext-\veff$.  
The latter gives a counter contribution that cancel
changes caused by the Coulomb interaction.

We now explain how to obtain an expression in practice.
Suppose that self-consistency has been somehow attained.
Then we have $\{\ei, \Psi_i\} \approx \{E_i, \Phi_i\}$
around \efermi. $\{\Psi_i\}$ is a complete set because
they come from some $\H0$, though the $\{\Phi_i\}$ are not.
Then we can expand 
$\Re[\Sigma(e_i)]|\Psi_i\rangle$ ($\approx \Re[\Sigma(E_i)]|\Phi_i\rangle$) 
in $\{\ei, \Psi_i\}$ as
\begin{eqnarray}
{\rm Re}[ \Sigma(\ei)]|\Psi_i\rangle  
= \sum_{j,i} |\Psi_j\rangle {\rm Re}[\Sigma(\ei)]_{ji}, \nonumber
\end{eqnarray}
where ${\rm Re}[\Sigma(\omega)]_{ij} = 
\langle \Psi_i|{\rm Re}[\Sigma(\omega)]|\Psi_j \rangle$.
Then we introduce an energy-independent operator ${ R}$ defined as
\begin{eqnarray}
{ R}=\sum_{j,i} |\Psi_j\rangle {\rm Re}[\Sigma(\ei)]_{ji} \langle\Psi_i|,
\nonumber
\end{eqnarray}
which satisfies ${ R} |\Psi_i\rangle = {\rm Re}[\Sigma(\ei)]|\Psi_i\rangle$.
Thus we can use this ${ R}$ instead of ${\rm Re}[\Sigma(E_i)]$
in \req{eq:qp};
however, ${ R}$ is not hermitian thus we take only
the hermitian part of ${ R}$ as $\vxc= {\rm Re}[{ R}]$;
\begin{widetext}
\begin{eqnarray}
\vxc = \frac{1}{2}\sum_{ij} |\Psi_i\rangle 
  \left\{ {{\rm Re}[\Sigma(\ei)]_{ij}+{\rm Re}[\Sigma(\ej)]_{ij}} \right\}
  \langle\Psi_j|,    \ \ \ \  {\rm mode\!-\!A}
\label{eq:veff}
\end{eqnarray}
\end{widetext}
for the calculation of $\{E_i, \Phi_i\}$ ($\approx \{\ei, \Psi_i\}$)
in \req{eq:qp}. Thus we have obtained a mapping 
$\veff \rightarrow \vgw(\omega) \rightarrow \veff$: for given $\veff$
we can calculate $\vxc$ in \req{eq:veff} through 
$\Sigma(\omega)$ in the $GW$A. With this $\vxc$ together with
$\vh$, which is calculated from the density for $G^0$ (or $\H0$), 
we have a new $\veff$.
The \qsgw\ cycle determines all $H^0, \veff$, $W$ and $G$ self-consistently.
As shown in Sec.~\ref{sec:result} and also in 
Refs.\ocite{faleev04,mark06qsgw,chantis06a}, \qsgw\ systematically
overestimates semiconductor band gaps a little, while the
dielectric constant $\epsilon_\infty$ is slightly too small
\cite{mark06qsgw}.

It is possible to derive \req{eq:veff} in a straightforward manner
from a norm-functional formalism. We first define a positive-definite norm functional 
$M(\veff) = \tr[ \Re[\DVo] \rho \Re[\DVo] ]$
to measure the size of pertubative contribution.
Here the weight function $\rho=\delta(\omega - \H0)$ defines the measure;
$\tr$ is for space, spin and  $\omega$. 
For fixed $\rho$, this $M(\veff)$ is treated as a functional of 
$\veff$ because $\veff$ determines $\DVo$ through \req{eq:dvo} in the $GW$A.
As $M(\veff) = \sum_{j,i} 
|\langle \Psi_i| \Re[\vgw(\ej)]- \veff| \Psi_j\rangle|^2$,
we can show its minimum occurs when \req{eq:veff} is satisfied
in a straightforward manner.
This minimization formalism clearly shows 
that \qsgw\ determines $\veff$ for a given $\vext$; 
in addition, it will be useful for formal discussions of conservation laws
and so on. 
The discussion in this paragraph is similar to that
given in Ref.~\onlinecite{mark06qsgw}, though
we use a slightly different $M(\veff)$.

\req{eq:veff} is derived from the requirement
so that $\{\ei, \Psi_i\} \approx \{E_i, \Phi_i\}$ around \efermi. 
This condition does not necessarily determine $\vxc$ uniquely.
It is instructive to evaluate how results change when
alternative ways are used to determine $\veff$.  In Ref.~\ocite{faleev04} 
we tested the following:
\begin{widetext}
\begin{eqnarray}
\vxc = 
      \sum_{i} |\psi_i\rangle {\rm Re}[\Sigma(\ei)]_{ii} \langle\psi_i|,
    + \sum_{i \ne j} |\psi_i\rangle 
       {\rm Re}[\Sigma(\efermi)]_{ij}
       \langle\psi_j|,   \ \ \ \  {\rm mode\!-\!B}
\label{eq:veffb}
\end{eqnarray}
\end{widetext}
In this form (which we denote as `mode-B'), the off-diagonal elements
are evaluated at \efermi.
The diagonal parts of \req{eq:veffb} and \req{eq:veff} are the same.
As noted in Ref.~\cite{faleev04}, and as
discussed in Sec.~\ref{sec:result}, Eqs.~(\ref{eq:veff}) and (\ref{eq:veffb}) yield
rather similar results, 
though we have found that mode-A results compare to experiment in
the most systematic way.

As the self-consistency through \req{eq:veff} (or \req{eq:veffb})
results in $\{\ei, \Psi_i\} \approx \{E_i, \Phi_i\}$,
we can attribute physical meaning to \bQP: 
we can use the \bQP\ in the independent-particle approximation \cite{ashcroft1976},
when, for example, modeling transport
within the Boltzmann-equation~\cite{fishetti88}.  It will
be possible to calculate scattering rates between \bQP\ given by
$\H0$, through calculation of various matrix elements (electron-electron,
electron-phonon, and so on).
The adiabatic connection path from $\hat{\H0}$ to $\hat{H}$ used in
\qsgw\ is better than the path in the Kohn-Sham theory
where the eigenfunction of $H_{\rm KS}$ (Kohn-Sham
Hamiltonian) evolves into the \dQP. 
Physical quantities along the path starting from $H_{\rm KS}$
may not be very stable. For example, the band gap can change 
very much along the path (it can change from metal to insulator
in some cases, e.g. in Ge and InN
\cite{mark06adeq}; \qsgw\ is free from this problem \cite{mark06qsgw}),
even if it keeps the density along the path.
[Note: Pines and Nozieres (Ref.~\cite{pines66}, Sec.~1.6) 
use the terms `bare QP' and `dressed QP' differently than what is meant here.
They refer to eigenstates of $\hat{H}$ as `bare QP,'
and spatially localized QP as `dressed QP' in the neutral Fermi liquid.]

From a theoretical point of view, 
the fully sc $GW$ \cite{eguiluz98,weiku02} looks reasonable because it is derived  
from the Luttinger-Ward functional $E[G]$. This apparently keeps the symmetry of $G$,
that is, $E[G]=E[{\cal R}[G]]$ where ${\cal R}[G]$ denotes 
some $G \to G$ mapping (any symmetry in Hamiltonian, e.g. time translation and gauge transformation);
this clearly results in the conservation laws for external perturbations 
\cite{baym61} because of 
Noether's theorem (exactly speaking,
we need to start from the effective action formalism 
for the dynamics of $G$ \cite{fukuda94}).
However, it contains serious problems in practice.
For example, fully sc $GW$ uses $W$ from $\Pi= -i G \times G$; 
this includes 
electron-hole excitations in its intermediate states with 
the weight of the product of renormalization factors $Z \times Z$. 
This is inconsistent with the expectation of
the Landau-Silin QP picture \cite{faleev04,Bechstedt97}.
In fact, as we discuss in Appendix~\ref{app:zfac}, the effects of $Z$
factor included in $G$ are 
well canceled because of the contribution from the vertex;
Bechstedt et al. showed the $Z$-factor cancellation by a practical calculation 
at the lowest order \cite{Bechstedt97}.
In principle, such a deficiency should be recovered 
by the inclusion of the contribution from the vertex;
however, we expect that such expansion series should be not efficient.

Generally speaking, perturbation theories in the dressed Green's 
function $G$ (as in Luttinger-Ward functional) can be very problematic 
because $G$ contains two different kinds of physical quantities to
intermediate states: the QP part (suppressed by the factor $Z$) and the 
incoherent part (e.g. plasmon-related satellites).
Including the sum of ladder diagrams into $\Pi$ via the Bethe-Salpeter
equation, should be a poorer approximation if $G$ is used instead of $G^0$, 
because the one-particle part is suppressed by $Z$
factors; also the contribution from the incoherent part can give physically
unclear contributions. The same can be said about the $T$-matrix
treatment \cite{springer98}.
Such methods have clear physical interpretation in a QP
framework, i.e. when the expansion is through $G^0$.
A similar problem is encountered in theories such as
``dynamical mean field theory''+$GW$ \cite{Biermann03}, 
where the local part of the proper polarization function is replaced
with a ``better'' function which is obtained 
with the Anderson impurity model.
This question, whether the perturbation should be based on $G$, or on
$G^0$, also appeared when Hedin obtained an equation to determine the
Landau QP parameters; See Eq. (26.12) in Ref.~\ocite{hedin65}.

As we will show in Sec.~\ref{sec:result} (see Ref.~\ocite{mark06qsgw} also),
\qsgw\ systematically overestimates band gaps,
consistent with systematic underestimation of $\epsilon_\infty$.
This looks reasonable because $W$ does not include the
electron-hole correlation within the RPA.  Its inclusion would effectively reduce
the pair excitation energy in its intermediate states.
If we do include such kind of correlation for $W$ at the level of
the Bethe-Salpeter equation, we will have an improved version of \qsgw.
However, the QPE obtained from $G^0 W$ with such a $W$ corresponds to
the $\Gamma=1$ approximation, from the perspective of the
$\Sigma = G^0 W \Gamma$ approximation, as used by Mahan and Sernelius \cite{mahan89};
the contribution from $\Gamma$ is neglected.
In order to include the contribution properly,
we need to use the self-energy derived from the functional derivative
of $E_c$ as shown in \req{ecform} in next section, where we need to include
the proper polarization $\Pi_\lambda$ which includes such Bethe-Salpeter
contributions; then we can include the corresponging $\Gamma$. 
It looks complicated, but it will be relatively easy to evaluate 
just the shift of QPE with neglecting the change of QPeig;
we just have to evaluate the change of $E_c$ numerically, 
when we add (or remove) an electron to $G_0$.
However, numerical evalution for these contributions
are demanding, and beyond the scope of this paper.


\subsection{Total energy}
\label{sec:totale}
Once $\veff$ is given, we can calculate the total energy 
based on the adiabatic connection formalism~\cite{fuchs02,miyake02,kotani98,fukuda94}.
Let us imagine an adiabatic connection path where the one-body Hamiltonian 
$\H0 = \frac{-\nabla^2}{2m} + \veff$
evolves into the total Hamiltonian $\hat{H}$, which is written as
\begin{eqnarray}
  &&\hat{H}= \hHk + \hVee +\hVext, \label{hami}\\
  &&\hHk =     \sum_\sigma \int d {\bf r} 
        \hat{\psi}^{\dagger}_{\sigma}({\bf r})
        ( - \frac{\nabla^2}{2m} )
        \hat{\psi}_{\sigma}({\bf r}), \\
  &&\hVext \!=\!\sum_{\sigma} \int d {\bf r} \vext_\sigma ({\bf r}) 
                              \hat{n}_\sigma ({\bf r} ), \label{eq:hvext}\\
  &&\hVee \!=\! \frac{1}{2}
          \sum_{\sigma\sigma'}
          \int \!\!d {\bf r} d {\bf r}' v(\bfr, \bfr') \times \nonumber \\
  &&         { \hat{\psi}^{\dagger}_{\sigma}({\bf r})
                   \hat{\psi}^{\dagger}_{\sigma'}({\bf r}')
                   \hat{\psi}_{\sigma'}({\bf r}')
                   \hat{\psi}_{\sigma}({\bf r}) }. 
\end{eqnarray}
$\hVeff$ is also defined with $\veff$ instead of $\vext$ in \req{eq:hvext}.
We use standard notation for the field operators
$\hat{\psi}_\sigma ({\bf r})$, spin index $\sigma$, and external potential
$\vext_{\sigma}({\bf r})$. 
We omit spin indexes below for simplicity. 

A path of adiabatic connection can be
parametrized by $\lambda$ as 
$\hat{H}^\lambda = \hat{H}^0 + \lambda(\hVext - \hVeff + \hVee)$.
Then the total energy $E$ is written as
\begin{widetext}
\begin{eqnarray}
E = E^0 + \int_0^1 d\lambda \frac{d E^\lambda}{d \lambda} 
= E^0 
+ \int_0^1 d\lambda 
\langle0_{\lambda}| \hVext - \hVeff |0_{\lambda}\rangle
+ \int_0^1 d\lambda 
\langle0_{\lambda}| \hVee |0_{\lambda}\rangle,
\label{eq:etot1}
\end{eqnarray}
\end{widetext}
where $|0_{\lambda}\rangle$ is the ground state for $\hat{H}^\lambda$.
We define $E^{\rm ext}=\int_0^1 d\lambda \langle0_{\lambda}| \hVext |0_{\lambda}\rangle$.
This path is different from the path used in DFT, where we take a path
starting from $\hat{H}_{\rm KS}$ to $\hat{H}$ while keeping the given density fixed.
Along the path of the adiabatic connection, the Green's
function changes from $G^0$ to $G$.  Because of our minimum-perturbation
construction, \req{eq:veff}, the QP parts (QPeig and QPE) 
contained in $G$ are well kept by $G^0$.
If $n_\lambda(\bfr)= \langle0_{\lambda}| \hat{n}({\bf r})|0_{\lambda}\rangle$
along the path is almost the same as $n_{\lambda=0}(\bfr)$, 
$E^0$ plus the second term in the RHS of \req{eq:etot1} is reduced to
$\langle0_{\lambda=0}| \hHk + \hVext |0_{\lambda=0}\rangle$
(this is used in $E^{\rm 1st}$ below).
The last term on the RHS of \req{eq:etot1} 
is given as $E^{\rm H} + E^{\rm x} + E^{\rm c}$, where
\begin{eqnarray}
&&E^{\rm H} = \frac{1}{2} \int_0^1 d\lambda 
n_\lambda(\bfr) v(\bfr,\bfr') n_\lambda(\bfr'), \label{eq:eh}\\
&&E^{\rm x} = -\frac{1}{2} \int_0^1 
d\lambda |n_\lambda(\bfr,\bfr')|^2 v(\bfr,\bfr'), \label{eq:ex}\\
&&E^{\rm c} = \frac{1}{2} \int_0^1 d\lambda
\langle0_{\lambda}| \hVee |0_{\lambda}\rangle -E^{\rm H} -E^{\rm x}
\end{eqnarray}
Here we used 
$n_\lambda(\bfr,\bfr')= \langle0_{\lambda}| \Psi^\dagger({\bf r}) 
\Psi({\bf r}')|0_{\lambda}\rangle$.

We define the 1st-order energy $E^{\rm{1st}}$ as the total energy 
neglecting $E^{\rm c}$:
%
\begin{eqnarray}
E^{\rm 1st} 
=  E_0^{\rm k} + E_0^{\rm ext} + E_0^{\rm H} + E_0^{\rm x},
\label{eq:e1st}
\end{eqnarray}
where subscript 0 means that
we use $n_{\lambda=0}(\bfr)$ instead of $n_{\lambda}(\bfr)$
(and same for $n_{\lambda}(\bfr,\bfr')$)
in the definition of $E^{\rm ext}$, $E^{\rm H}$ and $E^{\rm x}$;
$E_0^{\rm k}= \langle0_{\lambda=0}| \hHk|0_{\lambda=0}\rangle$.
This is the HF-like total energy, but with the QPeig given by $\veff$.

$E^{\rm c}$ is written as
\begin{eqnarray}
E^{\rm c}=\frac{1}{2}\int_0^1 d\lambda 
{\rm Tr}[v \Pi_\lambda (1-\lambda v \Pi_\lambda)^{-1} - v \Pi_\lambda],
\label{ecform}
\end{eqnarray}
where $\Pi_\lambda$ is the proper polarization function for the ground state of
$\hat{H}^\lambda$. The RPA makes the approximation
$\Pi_\lambda \approx \Pi_{\lambda=0}$ 
($\Pi_{\lambda=0}$ is simply expressed as $\Pi$ below).
The integral over $\lambda$ is then trivial, and
\begin{eqnarray}
&&E^{\rm c,RPA} = \frac{-1}{2}{\rm Tr}[  \log(1-v \Pi) + v \Pi], \nonumber \\
&&E^{\rm RPA}  = E^{\rm 1st} + E^{\rm c,RPA},
\label{erpac}
\end{eqnarray}
$E^{\rm RPA}$ denotes the RPA total energy.
$\Pi$ is given by the product of non-interacting Green's functions
$\Pi = -i G^0 \times G^0$, where $G^0$ is calculated from $\veff$. 
Thus we have obtained the total energy expression $E^{\rm RPA}$ for \qsgw.
As we have the smooth adiabatic connection 
from $\lambda=0$ to $\lambda=1$ in \qsgw (from \bQP\ to \dQP) 
as discussed in previous section,
we can expect that we will have better total energy than $E^{\rm RPA}$
where we use the KS eigenfunction and eigenvalues (where the band gap
can change much from \bQP\ to \dQP).
$E^{\rm RPA}$ will have characteristics
missing in the LDA, e.g. physical effects 
owing to charge fluctuations such as 
the van der Waals interaction, the mirror force on metal surfaces,
the activation energy, and so on.
However, the calculation of 
$E^{\rm c,RPA}$ is numerically very difficult,
because so many unoccupied states are needed.  Also deeper states can couple to
rather high-energy bands in the calculation of $\Pi$.  Few calculations have been carried out to
date~\cite{fuchs02,miyake02,ferdi02,marini06}. As far as we tested within
our implementation, avoiding systematic errors is rather difficult.
In principle, the expression $E^{\rm RPA}$ is basis-independent; however,
it is not so easy to avoid the dependence; for example, when we 
change the lattice constant in a solid, $E^{\rm c,RPA}$
artificially changes just because of the changes in the basis sets.
From the beginning, very high-level numerical accuracy for $E^{\rm RPA}$ required;
very slight changes of $E^{\rm c,RPA}$ results in non-negligible error
when the bonding originates from weak interactions such as the van der Waals interaction.
These are general problems in calculating the RPA-level of correlation energy,
even when evaluated from Kohn-Sham eigenfunctions.


\qsgw\ with \req{eq:veff} or \req{eq:veffb} 
can result in multiple self-consistent solutions for $G^0$ in some cases.
This situation can occur even in HF theory.
For any solution that satisfies the self-consistency as \req{eq:veff} or \req{eq:veffb},
we expect that it corresponds to some metastable solution.
Then it is natural to identify the lowest energy solution as
the ground state, that is, we introduce a new assumption that 
``the ground state is the solution with the lowest total energy among all solutions''.
In other words, the \qsgw\ method 
may be regarded as a construction that determines $\veff$
by minimizing $E^{\rm RPA}$ under the constraint of
\req{eq:veff} (or \req{eq:veffb}).
This discussion shows how \qsgw\ is connected to a variational principle.  The true ground state
is perturbatively constructed from the corresponding $\H0$.
However, total energy minimization is not necessary
in all cases, as shown in Sec.~\ref{sec:result}.
We obtain unique solutions (no multiple solutions) just with
\req{eq:veff} or \req{eq:veffb}
(Exactly speaking, we can not prove that multiple solutions do
not exist because we can not examine all the possibilities. However, we
made some checks to confirm that the results are not affected by initial conditions).
In the cases we studied so far, multiple solutions have been found, e.g.
in GdN, YH$_3$ and Ce \cite{chantis06fshell,sakuma_yh3}. These cases 
are related to the metal-insulator transition,
as we will detail elsewhere.
As a possibility, we can propose an extension
of \qsgw, namely to add a local static 
one-particle potential as a correction to \req{eq:veff}.
The potential is controlled to minimize $E^{\rm RPA}$. This is 
a kind of hybridization of \qsgw\ with 
the optimized effective potential method \cite{kotani98}.
See Appendix~\ref{app:etot} for further discussion as to why the
total energy minimization as functional of $\veff$ is not a suitable way to
determine $\veff$.

Finally, we discuss an inconsistency in the construction of the electron density
within the \qsgw\ method. The density used for the construction of
$\vh$ in the self-consistency cycle
is written as $n^{G^0}(\bfr) = \frac{-i}{2 \pi} \int d \omega G^0(\bfr,\bfr,\omega) e^{i \delta \omega}$,
which is the density of the non-interacting system with Hamiltonian $\H0$.
On the other hand, the density can be calculated
from $E^{\rm RPA}$ by the functional derivative with respect to $\vext$.
Since $E^{\rm RPA}$ is a functional of $\vext$,
we write it as $E^{\rm RPA}[\vext]$; its derivative 
gives the density $n^{E^{\rm RPA}}(\bfr) = \frac{\delta E^{\rm RPA}}{\delta \vext}$.
The difference in these two densities is given as
\begin{eqnarray}
&& n^{E^{\rm RPA}}(\bfr) - n^{G^0}(\bfr) \nonumber \\ 
&&\!\!=\! \frac{-i}{2 \pi} \int\! d1\! \int \!d2\! \left( \Sigma(1,2) - \vxc(1,2) \right)  
    \frac{\delta G^0(1,2)}{\delta \vext(\bfr)},
\label{eq:densityn}
\end{eqnarray}
where $\vxc$ is the static non-local potential defined in \req{eq:veff} or \req{eq:veffb}.
This difference indicates the size of inconsistency
in our treatment; from the view of the force theorem (Hellman-Feynman theorem),
we need to identify $n^{E^{\rm RPA}}(\bfr)$ as the true density, 
and $n^{G^0}(\bfr)$ for $\vh$ as the QP density.
We have not evaluated the difference yet.
\section{$GW$ methodological details}
\label{sec:method}
\subsection{Overview}
In the full-potential linear muffin-tin orbital method (FP-LMTO) and its
generalizations, eigenfunctions are expanded in linear
combinations of Bloch summed muffin-tin orbitals (MTO)
$\chi^{\bfk}_{\brl{j}}({\bf r})$  of wave vector $\bfk$ as
\begin{eqnarray}
\label{eq:lmtopsi}
\Psikn(\bfr) = \sum_{\brl{j}} z^{{\bf k}}_{\brl{j},n} \chi^{\bfk}_{\brl{j}}({\bf r}),
\end{eqnarray}
$n$ is the band index; $\Psikn(\bfr)$ is defined by the (eigenvector)
coefficients $z^{{\bf k}}_{\brl{j},n}$ and the shape of the
$\chi^{\bfk}_{\brl{j}}({\bf r})$.  The MTO we use here is a generalization
of the usual LMTO basis, and is detailed in
Refs.\ocite{mark06adeq,lmfchap}.  ${\bf{R}}$ identifies the site where the
MTO is centered within the primitive cell, $L$ identifies the angular momentum of the site.  There
can be multiple orbitals per $\brl$; these are labeled by $j$.  Inside a
MT centered at ${\bf{R}}$, the radial part of $\chi$ is spanned by radial
functions ($\varphi_{Rl}$, $\dot\varphi_{Rl}$ or $\varphi_{Rl}$,
$\dot\varphi_{Rl}$, $\varphi^z_{Rl}$) at that site.  Here $\varphi_{Rl}$ is
the solution of the radial Schr\"odinger equation at some energy
$\epsilon_\nu$ (usually, for $l$ channels with some occupancy, this is
chosen to be at the center of gravity for occupied states).
$\dot\varphi_{Rl}$ denotes the energy-derivative of $\varphi_{Rl}$;
$\varphi^z_{Rl}$ denotes local orbitals, which are solutions to the radial
wave equation at energies well above or well below $\epsilon_\nu$.  We
usually use two or three MTOs for each $l$ for valence electrons (we use
just one MTO for high $l$ channels with almost zero occupancy).  In any
case these radial functions are represented in a compact notation
$\{\varphi_{Ru}\}$.  $u$ is a compound index labeling $L$ and one of the
$(\varphi_{Rl}$, $\dot\varphi_{Rl}$, $\varphi^z_{Rl})$ triplet. The
interstitial is comprised of linear combinations of envelope functions
consisting of smooth Hankel functions, which can be expanded in terms of
plane waves~\cite{Bott98}.

Thus $\Psikn(\bfr)$ in \req{eq:lmtopsi}
can be written as a sum of augmentation and interstitial parts
\begin{eqnarray}
\Psikn(\bfr)
= \sum_{Ru}      \alpha^{{\bfk}n}_{Ru} \varphi^{\bf k}_{Ru}({\bf r})
 + \sum_{\bf G}  \beta^{{\bfk}n}_{\bf G} P^{\bf k}_{\bf G}({\bf r}),
\label{def:psiexp}
\end{eqnarray}
where the interstitial plane wave (IPW) is defined as
\begin{eqnarray}
P^{\bf k}_{\bf G}({\bf r}) =
\begin{cases}
 0                           & \text{if {\bf r}} \in \text{any MT} \\
\exp (i ({\bf k+G})\cdot{\bf r}) & \text{otherwise}
\end{cases}
\end{eqnarray}
and $\varphi^{\bf k}_{R u}$ are Bloch sums of $\varphi_{R u}$
\begin{eqnarray}
\varphi^{\bf k}_{R u}({\bf r}) &\equiv& \sum_{\bf T} \varphi_{R u}({\bf r-R-T}) \exp(i {\bf k\cdot{}T}).
\end{eqnarray}
{\bf T} and {\bf G} are lattice translation vectors in real and reciprocal
space, respectively. 
Eq.~(\ref{def:psiexp}) is equally valid in a LMTO or LAPW framework, and
eigenfunctions from both types of methods have been used in this $GW$
scheme \cite{usuda02,friedrich06}.  Here we restrict ourselves to
(generalized) LMTO basis functions, based on smooth Hankel functions.

Throughout this paper, we will designate eigenfunctions constructed from
MTOs as \VAL. Below them are the core eigenfunctions which we designate as
\CORE.  There are two fundamental distinctions between \val\ and \core:
first, the latter are constructed independently by integration of the
spherical part of the LDA potential, and they are not included in the
secular matrix. Second, the \core\ eigenfunctions are confined to MT
spheres~\cite{coretrunc}.  \core\ eigenfunctions are also expanded using
\req{def:psiexp} in a trivial manner ($\beta^{{\bfk}n}_{\bf G}=0$ and only
one of $\alpha^{{\bfk}n}_{Ru}$ is nonzero); thus the discussion below
applies to all eigenfunctions, \val\ and \core.  In order to obtain \CORE\
eigenfunctions, we calculate the LDA Kohn-Sham potential for the density
given by $\H0$, and then solve the radial Schr\"odinger equation.  In other
words, we substitute the nonlocal $\veff$ potential with its LDA counterpart
to calculate \CORE.  More details of the core treatment are given in
Sec.~\ref{sec:core}.

We need a basis set (referred to as the mixed basis) which
encompasses any product of eigenfunctions.  It is required for
the expansion of the Coulomb interaction $v$ (and also the
screened interaction $W$) because it connects the products as $\langle
\Psi \Psi | v |\Psi \Psi \rangle$.
Through Eq.~(\ref{def:psiexp}), products $\Psi_{{\bf k_1}n}\times\Psi_{{\bf
k_2}n'}$ can be expanded by $P^{\bf k_1+k_2}_{\bf G}({\bf r})$ in the
interstitial region because $P^{\bf k_1}_{\bf G_1}({\bf r}) \times P^{\bf
k_2}_{\bf G_2}({\bf r})= P^{\bf k_1+k_2}_{\bf G_1+G_2}({\bf r})$.  Within
sphere $R$, products of eigenfunctions can be expanded by
$B_{Rm}^{\bf{k_1+k_2}}({\bf r})$, which is the Bloch sum of the product
basis (PB) $\{B_{Rm}({\bf r})\}$, which in turn is constructed from the set
of products $\{\varphi_{Ru}({\bfr})\times\varphi_{Ru'}({\bfr})\}$.  For the
latter we adapted and improved the procedure of
Aryasetiawan~\cite{prodbasis94}.  
As detailed in Sec.~\ref{sec:mixed},
we define the mixed basis
$\{M^{\bf k}_I({\bf r}) \}\equiv \{ P^{\bf k}_{\bf G}({\bf r}), B_{Rm}^{\bf k}({\bf r})\}$,
where the index $I\equiv \{ {\bf G},Rm\}$ classifies the members of the basis.
By construction, $M^{\bf k}_I$ is a virtually complete 
basis, and efficient one for the expansion of $\Psikn$ products.
Complete information to generate the \gwa\ self-energy are
matrix elements $\langle \Psi_{{\bf q}n}| \Psi_{{\bf q-k}n'} M^{\bf k}_I \rangle$,
the eigenvalues $\varepsilon_{{\bf k}n}$,
the Coulomb matrix $v_{IJ}({\bf k})
\equiv \langle M^{\bf k}_{I} |v|  M^{\bf k}_{J} \rangle$,
and the overlap matrix $\langle M^{\bf k}_I |  M^{\bf k}_{J} \rangle$.
(The IPW overlap matrix is necessary because
$\langle P^{\bf k}_{\bf G} |  P^{\bf k}_{\bf G'} \rangle \ne 0$
for ${\bf G} \ne {\bf G}'$.)
The Coulomb interaction is expanded as
\begin{eqnarray}
\label{coulombexpand}
v({\bf r},{\bf r}') &=&
\sum_{{\bf k},I,J} |\tilde{M}^{\bf k}_I \rangle
v_{IJ}({\bf k}) \langle \tilde{M}^{\bf k}_J|,
\end{eqnarray}
where we define
\begin{eqnarray}
&& |\tilde{M}^{\bf k}_{I} \rangle \equiv \sum_{I'}
   |M^{\bf k}_{I'} \rangle (O^{\bf k})^{-1}_{I'I} \, , \\
&& O^{\bf k}_{I'I} = \langle M^{\bf k}_{I'} |  M^{\bf k}_I \rangle.
\end{eqnarray}
$W$ and the polarization function $\Pi$ shown below are expanded
in the same manner.

\begin{widetext}
The exchange part of $\Sigma$ is written in the mixed basis as
\begin{eqnarray}
\langle \Psiqn|\Sigma_{\rm x} |\Psiqm \rangle
&&=-\sum^{\rm BZ}_{{\bf k}}  \sum^{\rm  occ}_{n'}
\langle \Psiqn| \Psi_{{\bf q-k}n'} \tilde{M}^{\bf k}_I \rangle
v_{IJ}({\bf k})
\langle \tilde{M}^{\bf k}_J \Psi_{{\bf q-k}n'} | \Psiqm \rangle.
\label{eq:sigx}
\end{eqnarray}
It is necessary to treat carefully the
Brillouin zone (BZ) summation in \req{eq:sigx} and also \req{eq:sigc} 
because of the divergent character of $v_{IJ}(\bfk)$ at $\bfk \to 0$.
It is explained in Sec.~\ref{sec:kint}.

The screened Coulomb interaction $W_{IJ}({\bf q},\omega)$ is calculated
through Eq.~(\ref{eq:defw}), where the polarization function is written as
\begin{eqnarray}
\Pi_{IJ}({\bf q},\omega)
&&=
\sum^{\rm BZ}_{\bfk} \sum^{\rm occ}_{n \ispone} \sum^{\rm unocc}_{n'\isptwo}
\frac{
\langle \tilde{M}^{\bf q}_I \Psikn |\Psi_{{\bf q+k}n'} \rangle
\langle \Psi_{{\bf q+k}n'}| \Psikn \tilde{M}^{\bf q}_J \rangle
}{\omega-(\varepsilon_{{\bf q+k} n'\isptwo}-\varepsilon_{\bfk n\ispone})+i \delta} \nonumber\\
&&+ \sum^{\rm BZ}_{\bfk} \sum^{\rm  unocc}_{n \ispone} \sum^{\rm occ}_{n'\isptwo}
\frac{
\langle \tilde{M}^{\bf q}_I \Psi_{{\bf k}n} |\Psi_{{\bf q+k}n'} \rangle
\langle \Psi_{{\bf q+k}n'}| \Psi_{{\bf k}n} \tilde{M}^{\bf q}_J \rangle
}{-\omega-(\varepsilon_{\bfk n\ispone}-\varepsilon_{{\bf q+k} n'\isptwo})+i \delta}.
\label{eq:polf0}
\end{eqnarray}
When time-reversal symmetry is assumed, $\Pi$ can be simplified to read
\begin{eqnarray}
\Pi_{IJ}({\bf q},\omega)
&&=\sum^{\rm BZ}_{{\bf k}}  \sum^{\rm  occ}_{n} \sum^{\rm  unocc}_{n'}
\langle \tilde{M}^{\bf q}_I \Psi_{{\bf k}n} |\Psi_{{\bf q+k}n'} \rangle
\langle \Psi_{{\bf q+k}n'}| \Psi_{{\bf k}n} \tilde{M}^{\bf q}_J \rangle \nonumber \\
&& \times
\left(\frac{1}{\omega-\varepsilon_{{\bf q+k}n'}+\varepsilon_{{\bf k}n}+i \delta}
-\frac{1}{\omega+\varepsilon_{{\bf q+k}n'}-\varepsilon_{{\bf k}n}-i \delta}\right). \label{dieele}
\label{eq:polf}
\end{eqnarray}
We developed two kinds of tetrahedron method for the Brillouin zone (BZ)
summation entering into $\Pi$.  One follows the technique of Rath and
Freeman~\cite{rath75}.  The other, which we now mainly use, first
calculates the imaginary part (more precisely the anti-hermitian part) of
$\Pi$, and determines the real part via a Hilbert transformation
(Kramers-Kr\"onig relation); see Sec.~\ref{sec:tetra}.
The Hilbert transformation approach significantly reduces the computational
time needed to calculate $\Pi$ when a wide range of $\omega$ is needed.  A
similar method was developed by Miyake and Aryasetiawan~\cite{miyake00}.

The correlation part of $\Sigma$ is
\begin{eqnarray}
\langle \Psiqn |\Sigma_{\rm{c}}(\omega) |\Psiqm \rangle
&& = \sum^{\rm BZ}_{\bf k}  \sum^{\rm  All}_{n'} \sum_{IJ}
\langle \Psiqn| \Psi_{{\bf q-k}n'} \tilde{M}^{\bf k}_I \rangle
\langle \tilde{M}^{\bf k}_J \Psi_{{\bf q-k}n'} | \Psiqm \rangle  \nonumber \\
&& \times \int_{-\infty}^{\infty} \frac{i d\omega'}{2 \pi}
W^{\rm{c}}_{IJ}({\bf k},\omega')
\frac{1}{-\omega'+\omega-\varepsilon_{{\bf q-k}n'} \pm i \delta}.
\label{eq:sigc}
\end{eqnarray}
\end{widetext}
where $W^{\rm{c}} \equiv W - v$
($-i \delta$ must be used for occupied states, $+i \delta$ for unoccupied states).
Sec.~\ref{sec:sigcint} explains how the $\omega$-integration is performed.

\subsection{Core treatment}
\label{sec:core}
Contributions from core (or semi-core) eigenfunctions require special cares.
In our $GW$, \CORE\ is divided into groups, \CORE1 and \CORE2.
Further, \val\ can be divided into ``core'' and ``val''.
Thus all eigenfunctions are divided into the following groups:
\begin{equation}
\quad \pstree[nodesep=2pt,levelsep=20pt]{\TR{\rm All \ eigenfunctions}}
  {\pstree{\TR{\core}}
     {\pstree{\TR{\core1}}{}
      \pstree{\TR{\core2}}{}
    }
   \pstree{\TR{\val}}
     {\pstree{\TR{\rm ``core''}}{}
      \pstree{\TR{\rm ``val''. }}{}
    }
}
\label{cvmap}
\end{equation}

\VAL\ states are
computed by the diagonalization of a secular matrix for MTOs;
thus they are completely orthogonal to each other.
\val\ can contain core eigenfunctions we denote as ``core''.
For example, we can treat the Si 2$p$ core as ``core''.
Such states are reliably determined by using local
orbitals, tailored to these states~\cite{mark06adeq}.  

\CORE1 is for deep core eigenfunctions.  Their screening is small,
and thus can be treated as exchange-only core.
The deep cores are rigid with little freedom to be deformed;
in addition, CORE2+VAL is not included in these cores. 
Thus we expect they give little contribution to $\Pi$
and to $\Sigma_{\rm c}$ for CORE2+VAL.
Based on the division of \CORE\ according to \req{cvmap}, 
we evaluate $\Sigma$ as 
\begin{eqnarray}
\Sigma=   \Sigma^{\rm CORE1}_{\rm x} 
        + \Sigma^{\rm CORE2+VAL}_{\rm x}
        + \Sigma_{\rm c}^{\rm CORE2+VAL}.
\label{cvdivide}
\end{eqnarray}
(We only calculate the matrix elements $\langle \Psi_i |\Sigma|\Psi_j \rangle$,
 where $i$ and $j$ belongs to CORE2+VAL, not to CORE1.)
We need to generate two kinds of PB; one for $\Sigma^{\rm CORE1}_{\rm x}$, the
other for $\Sigma^{\rm CORE2+VAL}_{\rm x}$ and $\Sigma^{\rm CORE2+VAL}_{\rm c}$.
As explained in Sec.~\ref{sec:mixed}, these PB
should be chosen taking into account what combination of eigenfunction
products are important.  States {\rm CORE2+VAL} are usually included in
$\Pi$, which determines $W$.  Core eigenfunctions sufficiently deep (more
than $\sim$ 2 Ry below $\EF$), are well-localized within their MT spheres.
For such core eigenfunctions, we confirmed that results are little affected
by the kind of core treatments (\CORE1, \CORE2, 
and ``core'' are essentially
equivalent); see Ref.~\ocite{mark06adeq}.

As concerns their inclusion in the generation of $\Pi$ and $\Sigma$,
\req{cvdivide} means that not only \VAL\ but also \CORE2 are treated on the
same footing as ``val''.  However, we have found that it is not always
possible to reliably treat shallow cores (within $\sim$ 2 Ry below $\EF$)
as \CORE2.  Because \CORE\  
eigenfunctions are solved separately, the orthogonality
to \VAL\ is not perfect; this results in a small but uncontrollable error.
The nonorthogonality problem is clearly seen in $v \Pi({\bf q},\omega)$ as
$q\to 0$: cancellation between denominator and numerator becomes
imperfect.  (We also implemented a procedure that enforced
orthogonalization to \VAL\ states, but it would sometimes produce unphysical shapes
in the core eigenfunctions.)  Even in LDA calculations, MT spheres can be
often too small to fully contain a shallow core's eigenfunction.  Thus we
now usually do not use \CORE2; for such shallow cores, we usually treat it
as ``core'' $\in$ \VAL; or as \CORE1 when they are deep enough.  We have
carefully checked and confirmed the size of contributions from cores by
changing such grouping, and also intentional cutoff of the core
contribution to $W$ and so on; see Ref.~\ocite{mark06adeq}.

\subsection{Mixed basis for the expansion of $\Psi \times \Psi$}
\label{sec:mixed}

A unique feature of our $GW$ implementation is its mixed basis set.
This basis, which is virtually complete for the expansion of the products
$\Psi\times\Psi$, is central for the efficient expansion of products of
relatively localized functions, and essential for proper treatment of very
localized states such as core states or $f$ systems.  Products within a MT
sphere are expanded by the PB procedure originally developed by
Aryasetiawan \cite{prodbasis94}.  We use an improved version explained
here.  For the PB construction we start from the set of radial functions
$\{\phi_{ls}(r)\}$, which are used for the augmentation for $\Psi$ in a MT
site.  $l$ is the principal angular momentum, $s$ is the other index
(e.g. we need $s=1$ for $\phi$, and $s=2$ for $\dot{\phi}$ in addition to
$s=3,4,..$ for local orbitals and cores).  
The products $\phi_{ls}(r) \times \phi_{l's'}(r)$
can be re-ordered by the total angular momentum $l_{\rm
p}=l+l',...,|l-l'|$.  Then the possible products of the radial functions are 
arranged by $l_{\rm p}$.  
To make the computation efficient, we need to reduce the 
dimension of the radial products as follows:
\begin{itemize}
\item[(1)] 
Restrict the choice of possible combinations $\phi_{ls}(r)$ and
$\phi_{l's'}(r)$.  In the calculation of $W$, one $\phi$ is used for
occupied states, the other for unoccupied states~\cite{prodbasis94}.
In the calculation of $\langle \Psiknp |G\times W |\Psikn \rangle$,
$\Psikmstar \times \Psikn $ appears, with $\Psikmstar$ coming from $G$.
Thus all possible products can appear; however, we expect the important
contributions come from low energy parts.  Thus, we define two sets
$\Phi_{\rm OCC}$ and $\Phi_{\rm UNOCC}$ as the subset of
$\{\phi_{ls}(r)\}$.  $\Phi_{\rm OCC}$ includes $\phi_{ls}(r)$ mainly for
occupied states (or a little larger sets), and $\Phi_{\rm UNOCC}$ is
$\Phi_{\rm OCC}$ plus some $\phi_{ls}(r)$ for unoccupied states (thus
$\Phi_{\rm OCC} \in \Phi_{\rm UNOCC}$). Then we take all possible products
of $\phi_{ls}(r) \times \phi_{l's'}(r)$ for $\phi_{ls}(r) \in \Phi_{\rm
OCC}$ and $\phi_{l's'}(r) \in \Phi_{\rm UNOCC}$.
Following Aryasetiawan~\cite{prodbasis94}, we usually do not include
$\dot{\phi}$-kinds of radial functions in these sets (we have checked in a
number of cases that their inclusion contributes little).

\item[(2)] Restrict $l_{\rm p}$ to be less than some cutoff $l^{\rm MAX}_{\rm p}$.
removing expensive product basis with high $l_{\rm p}$.
In our experience, we need $l^{\rm MAX}_{\rm p}= 2\times$ 
(maximum $l$ with non-zero (or not too small) electron occupancy)
is sufficient to predict band gaps to $\sim0.1$eV,
e.g. we need to take $l^{\rm MAX}_{\rm p}=4$ for transition metal atoms.

\item[(3)] Reduce linear dependency in the radial product basis.  For each
$l_{\rm p}$, we have several radial product functions.  We calculate the
overlap matrix, make orthogonalized radial functions from them, and
omit the subspace whose overlap eigenvalues are smaller than some specified
tolerance. The tolerance for each $l_{\rm p}$ can be different, and
typically tolerances for higher $l_{\rm p}$ can be coarser than for lower
$l_{\rm p}$.
\end{itemize}
This procedure yields a product basis, $\sim 100$ to $150$ functions for a
transition metal atom, and less for simple atoms (see
Sec.~\ref{sec:result_mixed} for GaAs).  

There are two kinds of cutoffs in the IPW part of the mixed basis: $|{\bf q
+G}|^\Psi_{\rm Max}$ for eigenfunctions \req{def:psiexp}, and $|{\bf q
+G}|^W_{\rm Max}$ for the mixed basis in the expansion of $W$. 
In principle, $|{\bf q
+G}|^W_{\rm Max}$ must be $2\times|{\bf q +G}|^\Psi_{\rm Max}$ to span the
Hilbert space of products.  However, it is too expensive.  The
computational time is strongly controlled by the size of the mixed basis.
Thus we usually take small $|{\bf q +G}|^W_{\rm Max}$, rather smaller than
$|{\bf q +G}|^\Psi_{\rm Max}$ (the computational time is much less strongly
controlled by $|{\bf q +G}|^\Psi_{\rm Max}$).  As we illustrate in
Sec.~\ref{sec:result_mixed}, $\sim$0.1~eV level accuracy can be realized
for cutoffs substantially below $2\times|{\bf q +G}|^\Psi_{\rm Max}$.

For the exchange part of \CORE1, we need to construct
another PB. It should include products of \CORE1 and VAL.  We construct it from
$\phi_{\rm CORE1} \times \psi_{\rm VAL}$, where $\psi_{\rm VAL} \in \Phi_{\rm UNOCC}$,
so as to make it safer ($\Phi_{\rm UNOCC}$ is bigger than $\Phi_{\rm OCC}$).

We also have tested other kinds of mixed basis which are smoothly
augmented; we augment IPWs and construct smoothed product basis (value and
slope vanishing at the MT boundary).  However, little computational
advantage was realized for such a mixed basis.

\subsection{Tetrahedron method for $W$}
\label{sec:tetra}

\req{eq:polf0} requires an evaluation of this type of BZ integral;
\begin{eqnarray}
X(\omega) = \sum_\bfk^{\rm BZ} T_\bfk
\frac{f(\eak) [1-f(\ebk)]}{\omega - (\ebk-\eak) \pm i \delta },
\end{eqnarray}
where $T_\bfk$ is a matrix element and $f(\varepsilon)$ is the Fermi function.
To evaluate this integral in the tetrahedron method, we divide the BZ into
tetrahedra.  $T_\bfk$ is replaced with
its average at the four corners of $\bfk$, $\overline{T_\bfk}$.
We evaluate the integral within a tetrahedron, linearly interpolating $\eak$
and $\ebk$ between the four corners of the tetrahedron.  In the metal 
case, we have to divide the tetrahedra into smaller tetrahedra; in each
of them, $f(\eak) [1-f(\ebk)] = 1$ or $=0$ are satisfied; see
Ref.~\ocite{rath75}.  Thus we only pick up the smaller tetrahedron $\iDelta
\Omega$ satisfying $f(\eak) [1-f(\ebk)] = 1$ and calculate
\begin{widetext}
\begin{eqnarray}
\frac{1}{\pi}{\rm Im} \iDelta X(\omega) = \overline{T_\bfk} \int_{{\it \Delta} \Omega} d^3k
\delta( \omega - (\ebk-\eak) ),
\end{eqnarray}
Based on the assumption of linear interpolation,
the integral in ${\rm Im}\iDelta X(\omega)$ equals the
area of the cross section of tetrahedron in a plane specified 
by energy $\omega = \ebk-\eak$.
We create a histogram of energy windows $[\omega(i), \omega(i+1)], i=0,1,...$,
by calculating the weight falling in each window as
$\int_{\omega(i)}^{\omega(i+1)} {\rm Im} \iDelta X(\omega) d\omega$.
We take windows specified as $\omega(i) = a \ i + b \ i^2 \ (i\!=\!0,1,..)$,
where we typically take $a \sim 0.05$eV and $b\sim 1-10$~eV.
Summing over contributions from all tetrahedra, we finally have 
\begin{eqnarray}
{\rm Im}\,X([\omega(i), \omega(i+1)])
= \int_{\omega(i)}^{\omega(i+1)} d\omega\, {\rm Im}X(\omega) 
=\sum_\bfk w_\bfk(i)\,T_\bfk.
\end{eqnarray}
Applying this scheme to \req{eq:polf}, we have
\begin{eqnarray}
{\rm Im}\Pi(q,[\omega(i), \omega(i+1)])
= \sum_\bfk \sum_{n,n'} w^{\bfq nn'}_\bfk(i) 
T^{\bfq nn'IJ}_\bfk,
\label{eq:tetimpi}
\end{eqnarray}
where $T^{\bfq nn'IJ}_\bfk= \langle \tilde{M}^{\bf q}_I \Psi_{{\bf k}n}
|\Psi_{{\bf q+k}n'} \rangle \langle \Psi_{{\bf q+k}n'}| \Psi_{{\bf k}n}
\tilde{M}^{\bf q}_J \rangle$.  The real part of $\Pi$ is calculated through
a Hilbert transform of ${\rm Im}\,\Pi$ (Kramers-Kr\"onig relation).
\end{widetext}

Some further considerations are as follows.
\setcounter{Alist}{0}
\begin{list}{({\rm \Alph{Alist}})\,}{\leftmargin 0pt \itemindent 24pt \usecounter{Alist}\addtocounter{Alist}{0}}
\item 
 For band index $n$, $\varepsilon_{\bfk n}$ may be degenerate.
 When it occurs, we merely symmetrize $w^{\bfq nn'}_\bfk(i)$ with
 respect to the degenerate $\varepsilon_{\bfk n}$.
\item 
 When \req{eq:polf0} is evaluated without time-reversal symmetry assumed,
 windows for negative energy must be included because ${\rm
 Im}\,\Pi(q,-\omega) \ne {\rm Im}\,\Pi(q,\omega)$.
\item 
 In some limited tests, we found that linearly interpolating $T_\bfk$
 within the tetrahedron, instead of using $\overline{T_\bfk}$, did little
 to improve convergence.
\item 
 We sometimes use a ``multi-divided'' tetrahedron scheme to improve on the
 resolution of the energy denominator.  We take a doubled $\bfk$-mesh
 when generating $w^{\bfq nn'}_\bfk(i)$.  For example, we calculate
 $\Pi$ with a $4\!\times\!4\!\times\!4$ $\bfk$-mesh for $k$ sum in
 \req{eq:tetimpi}; but we use a $8\!\times\!8\!\times\!8$ mesh when we
 calculate $w^{\bfq nn'}_\bfk(i)$.  Then the
 improvement of $\Pi$ is typically intermediate between the two meshes: we obtain
 results between the $4\!\times\!4\!\times\!4$ and
 $8\!\times\!8\!\times\!8$ results in the original method.
\end{list}

\subsection{Brillouin-zone integral for the self-energy; 
the smearing method and the offset-$\Gamma$ method.}
\label{sec:kint}

\subsubsection{Smearing method}
\label{sec:smear}

To calculate $\Sigma_{\rm x}$
and $\Sigma_{\rm c}$, Eqs.~(\ref{eq:sigx}) and (\ref{eq:sigc}),
each pole at $\varepsilon_{{\bf q}-{\bf k} n'}$
is slightly smeared.
The imaginary part, proportional to $\delta(\omega -\varepsilon_{{\bf q- k} n'})$,
is broadened by a smeared function $\bar{\delta}(\omega -\varepsilon_{{\bf q- k} n'})$.
In order to treat metals, this smearing procedure is necessary.
Usually we use a Gaussian for the smeared function
\begin{eqnarray}
&&\bar{\delta}(\omega) = \frac{1}{\sqrt{2 \pi} \sigma} \exp( -\frac{\omega^2}{2 \sigma^2}),
\label{eq:smeargauss}
\end{eqnarray}
though other forms have been tested as well.

We explain the smearing procedure by illustrating it for $\Sigma_{\rm x}$.
\req{eq:sigx} becomes
\begin{widetext}
\begin{eqnarray}
&&\langle \Psiqn|\Sigma_{\rm x} |\Psiqm \rangle
= \sum_{\bf k}^{\rm BZ} \rho_{{\bf q}nm}({\bf k}), \label{eq:sxqn}\\
&&\rho_{{\bf q}nm}({\bf k}) \equiv 
\sum^{\rm  all}_{n'} \sum_{IJ}
\bar{\theta}(E_{\rm F} -\varepsilon_{{\bf q- k} n'})
\langle \Psiqn| \Psi_{{\bf q-k}n'} \tilde{M}^{\bf k}_I \rangle v_{IJ}({\bf k})
\langle \tilde{M}^{\bf k}_J \Psi_{{\bf q-k}n'} | \Psiqm \rangle
\label{sxsmear2}
\end{eqnarray}
\end{widetext}
where $\bar{\theta}(\omega)$ is 
a smeared step function,
${d \bar{\theta}(\omega)}/{d \omega}= \bar{\delta}(\omega)$.

Owing to the sudden Fermi-energy cutoff in the metals case,
$\rho_{{\bf q}nm}({\bf k})$ may not vary smoothly with $\bfk$.  Increasing
$\sigma$ smooths out $\rho_{{\bf q}nm}({\bf k})$, making it more rapidly
convergent in spacing between $\bfk$-points, at the expense of introducing
a systematic error to the fully $\bfk$-converged result.  With a denser
$\bfk$ mesh, smaller $\sigma$ can be used, which reduces the systematic
error.  In practice we can obtain converged results for given $\sigma$ with
respect to the number of $\bfk$ points, and then take the $\sigma\to 0$
limit.


\subsubsection{Offset-$\Gamma$ method}
\label{sec:offsetgamma}

The integrand in \req{eq:sxqn} contains divergent term proportional to
$1/|{\bf k}|^2$ for ${\bf k} \to 0$.  In order to handle this divergence we
invented the offset-$\Gamma$ method.  It was originally developed by
ourselves~\cite{kotani02} and is now used by other
groups~\cite{alouani03,yamasaki03}; it is numerically essentially equivalent
to the method of Gygi and Baldereschi~\cite{Gygi86}, where
the divergent part is treated analytically.

We begin with the method of Gygi and Baldereschi~\cite{Gygi86}.  The Coulomb
interaction $v_{IJ}({\bf k})$, which is a periodic function in $\bfk$,
includes a divergent part $v^0_{IJ}(\bfk) \propto U^{0}_I(\bfk)
U^{0*}_J(\bfk) F(\bfk)$ where $F(\bfk)\to 1/|\bfk|^2$ as $\bfk \to 0$.
$U^{0}_J(\bfk)$ are coefficients to plane wave $e^{i \bfk \bfr}$ in the
mixed basis expansion.  They divided the integrand into two
terms, one with no singular part which is treated
numerically; the other is a combination of analytic functions that contain
the singular part:
\begin{widetext}
\begin{eqnarray}
\langle \Psiqn|\Sigma_{\rm x} |\Psiqm \rangle
&=& \sum_{\bf k}^{\rm BZ} (\rho_{{\bf q}nm}({\bf k}) - \rho^0_{{\bf q}nm}({\bf k}))
 + A_{\bfq nm}\sum_{\bf k}^{\rm BZ}  F({\bf k}), \label{sxdecomp}\\
\text{where}\nonumber\\
\rho^0_{{\bf q}nm}({\bf k}) &=& A_{\bfq nm} F({\bf k})\\
A_{\bfq nm} &=& \lim_{\bfk \to 0} |\bfk|^2 \rho_{{\bf q}nm}({\bf k}).
\end{eqnarray}
\end{widetext}
As for the first term on the right-hand side (RHS) of Eq.~(\ref{sxdecomp}), 
the integrand $\rho_{{\bf q}nm}({\bf k}) - \rho^0_{{\bf q}nm}({\bf k})$
is a smooth function in the BZ with no singularity 
(more precisely, it can contain a part $\propto k_x/|\bfk|^2$,
however, it adds zero contribution around $\bfk=0$ because it
is odd in ${\bf k}$); it is thus easily evaluated numerically.  The second term is
analytically evaluated because $F(\bfk)$ is chosen 
to be an analytic function as shown below.
We evaluate the first term by numerical integration on a discrete mesh in a
primitive cell in the BZ.  The mesh is given as
\begin{eqnarray}
{\bf k}(i_1,i_2,i_3) &=& 2 \pi (\frac{i_1}{N_1} {\bf b}_1 
+ \frac{i_2}{N_2} {\bf b}_2 + \frac{i_3}{N_3} {\bf b}_3),
\label{kmesh}
\\
\sum_{\bf k}^{\rm BZ} &\approx& \frac{1}{N_1N_2N_3} \sum_{i_1,i_2,i_3}.
\nonumber
\end{eqnarray}

Forms of the analytic functions $F({\bf k})$ are chosen so that it can be
analytically integrated.
We choose $F({\bf k})$ as
\begin{eqnarray}
{F}({\bf k})=\sum^{\rm All}_{\bf G} \frac{\exp(-\alpha |{\bf q + G}|^2) }{|{\bf q + G}|^2 },
\label{auxf}
\end{eqnarray}
where ${\bf G}$ denotes all the inverse reciprocal vectors and $\alpha$ is
a parameter.  ${F}({\bf k})$ is positive definite, periodic in BZ and has
the requisite divergence at ${\bf k} \to 0$.  (Gygi and Baldereschi used
a different function in Ref.~\cite{Gygi86}, but it satisfies the same
conditions.)
We can easily evaluate ${F}({\bf{k}})$ analytically.  Thus it is possible
to calculate $\Sigma_{\rm x}$ if we can obtain the coefficients $A_{\bfq
nm}$.  However, it is not easy to calculate $A_{\bfq nm}$.  This is
especially true for $\Sigma_{\rm c}$, \req{eq:sigc}, where the coefficients
are energy-dependent.

The offset-$\Gamma$ method avoids explicit evaluation of
$A_{\bfq nm}$, while retaining accuracy essentially equivalent to the method
described above.  It evaluates the ${\bf{k}}$-integral in \req{eq:sxqn}
as a discrete sum
\begin{eqnarray}
\sum_{\bf k}^{\rm BZ} \approx \frac{1}{N_1N_2N_3} \widetilde{\sum_{i_1,i_2,i_3}} \label{newmesh},
\end{eqnarray}
where $\widetilde{\sum}$ denotes the sum for ${\bf k}(i_1,i_2,i_3)$ but 
$\bfk=0$ is replaced by the offset-$\Gamma$ point ${\bf Q}=(Q,0,0)$.
${\bf Q}$ is near $\bfk=0$, and chosen so as to integrate
${F}({\bf{k}})$ exactly:
\begin{eqnarray}
\sum_{\bf k}^{\rm BZ} {F}({\bf k}) =
\frac{1}{N_1N_2N_3} \widetilde{\sum_{i_1,i_2,i_3}} {F}({\bf k}).
\label{chooseq}
\end{eqnarray}
Then \req{eq:sxqn} becomes
\begin{widetext}
\begin{eqnarray}
&&\langle \Psiqn |\Sigma_{\rm x} |\Psiqm \rangle
\approx \frac{1}{N_1N_2N_3} \widetilde{\sum_{i_1,i_2,i_3}} \rho_{{\bf q}nm}(\bfk)  \nonumber\\
&&= \frac{1}{N_1N_2N_3} \widetilde{\sum_{i_1,i_2,i_3}}(\rho_{{\bf q}nm}(\bfk)-\rho^0_{\bfq nm}(\bfk)) 
 +\frac{A_{\bfq nm}}{N_1N_2N_3} \widetilde{\sum_{i_1,i_2,i_3}} F(\bfk). \label{offg1}
\end{eqnarray}
\end{widetext}
For larger $N_1N_2N_3$, $(Q,0,0)$ is closer to $(0,0,0)$,
thus the first term on the RHS of \req{offg1} is little different from
the sum with the mesh \req{kmesh}.
Then the second term in \req{offg1} is exactly the same as
the second term in \req{sxdecomp} because of \req{chooseq}.
Thus the simple sum $\displaystyle \frac{1}{N_1N_2N_3} \widetilde{\sum_{i_1,i_2,i_3}}$
can reproduce the results given by the method \req{sxdecomp}.

In practical applications, we have to take some set of ${\bf Q}$ points to
preserve the crystal symmetry.  Typically we prepare six ${\bf Q}$ points,
$(\pm Q,0,0),(0,\pm Q,0),(0,0,\pm Q)$, and then generate all possible
points by the crystal symmetry. The weight for each $\bfQ$ should be
$1/N_1N_2N_3$ divided by the total number of ${\bf Q}$ points. The value of
$Q$ is chosen to satisfy \req{chooseq}.  It evidently depends on the
choice of ${F}({\bf{k}})$; in particular when ${F}$ is given by
Eq.~(\ref{auxf}), ${\bf Q}$ depends on $\alpha$.  A reasonable choice for
$\alpha$ is $\alpha \to 0$ (then no external scale is introduced).
However, we found $\alpha=1.0$ a.u. is small enough for simple solids, and
the results depend little on whether $\alpha=1.0$ or $\alpha\to{0}$.

In addition, we make the following approximation:
\begin{widetext}
\begin{eqnarray}
&&\rho_{{\bf q}n}({\bf Q}) \approx \sum^{\rm  all}_{n'} \sum_{IJ}
\bar{\theta}(E_{\rm F}-\varepsilon_{{\bf q} n'})
\langle \Psi_{{\bf q}n}| \Psi_{{\bf q}n'} \tilde{M}^{\bf k=0}_I \rangle v_{IJ}({\bf Q})
\langle \tilde{M}^{\bf k=0}_J \Psi_{{\bf q}n'} | \Psi_{{\bf q}n} \rangle.
\end{eqnarray}
\end{widetext}
That is, the matrix element is not evaluated at $\bfk=\bfQ$ but at
$\bfk=0$.  This is not necessary, but it reduces the computational costs
and omits the contribution $\propto k_x/|\bfk|^2$ which gives no
contribution around $\bfk=0$.

Finally, we use crystal symmetry to evaluate \req{sxsmear2}: $v_{IJ}(\bfk)$
and also $W^{\rm c}_{IJ}(\bfk)$ are calculated only at the irreducible
$\bfk$ points and at the inequivalent offset-$\Gamma$ points.


We also tested a modified version of the offset-$\Gamma$
method with another kind of BZ mesh (it is not used for any results in this paper).
The uniform mesh is shifted from the $\Gamma$ point:
\begin{eqnarray}
{\bf k}(i_1,i_2,i_3)= 2 \pi \left(\frac{i_1+\frac{1}{2}}{N_1} {\bf b}_1 
+ \frac{i_2+\frac{1}{2}}{N_2} {\bf b}_2 + \frac{i_3+\frac{1}{2}}{N_3} {\bf b}_3 \right).
\label{kmesh2}
\end{eqnarray}
Then we evaluate the BZ integral as
\begin{eqnarray}
\sum_\bfk^{\rm BZ} F(\bfk) \to \sum_{i_1,i_2,i_3} W_\bfk F(\bfk)
+ \frac{W_\bfQ}{N_1 N_2 N_3} F(\bfQ).
\label{newmesh2}
\end{eqnarray}
where $\bfk(i1,i2,i3)$ is used.  $W_\bfQ$ is a parameter given by hand to
specify the weight for the offset-$\Gamma$ point $\bfQ$.
(In principle, $\frac{W_\bfQ}{N_1 N_2 N_3} F(\bfQ)$ must give no contribution
 as $N_1, N_2, N_3 \to \infty$. Thus \req{newmesh2} can be taken a trick to accelarate
 the convergence on ${N_1 N_2 N_3}$; this is necessary in practice).
We usually use a small value, e.g., $W_\bfQ\sim0.01$ or less, 
but taken not too small so that it does not cause numerical problems.
Integration weights are $W_\bfk= 1/(N_1 N_2 N_3)$
except for those closest to $\Gamma$.  These latter
$W_{({\rm shortest} \ \bfk)}$ are chosen
so that the sum of all $W_\bfk$ and $W_\bfQ$ is unity.
Then $\bfQ$ is determined in the same manner in
the original offset-$\Gamma$ method.
This scheme picks up the divergent part of 
integral correctly, and can advantageous in some cases, in particular
for oddly shaped Brillouin zones.

The original method with \req{kmesh}
has difficulty in treating anisotropic systems like a
one-dimensional atomic chain.  In such a case, we can not determine
reasonable $\bfQ$ for the BZ division for, e.g., ($N_1=N_2=1$ and $N_3=$
large number), while the modified form \req{kmesh2} has no
difficulty.  We can choose $\bfQ$ close to $\Gamma$ (any $\bfQ$ can be
chosen if it is close enough to $\Gamma$).  As $W_\bfQ$ becomes smaller, 
so does $\bfQ$.  
Two final points relevant to the modified version are:
\begin{itemize}
\item In some anisotropic cases (e.g. anti-ferromagnetic II NiO), we
need to use negative $W_\bfQ$ because the shortest $\bfk$ on regular mesh
is already too short and the integral of $F(\bfq)$ evaluated on the 
mesh of Eq.~(\ref{kmesh2}) already exceeds the exact value.  However, the
modified version works even in such a case.
\item In some cases (e.g. Si), the shifted mesh can be somewhat
disadvantageous because certain symmetry operations can map some mesh points
to new points not within the shifted mesh, \req{kmesh2}.
Then the QP energies that are supposed to be degenerate
are not precisely so for numerical reasons.
One solution is to take denser $\bfk$ mesh to avoid the effect 
of discretization. Our current implementation allows us to 
compute $\Pi$ and $\Sigma$ with different meshes, \req{kmesh} or \req{kmesh2}.
This is sometimes advantageous to check the stability 
of calculations with respect to the ${\bf{k}}$ mesh.
\end{itemize}

\subsection{$\omega'$-integral for $\Sigma^{\rm c}$}
\label{sec:sigcint}

\begin{figure}
\begin{center}
\includegraphics[scale=0.5]{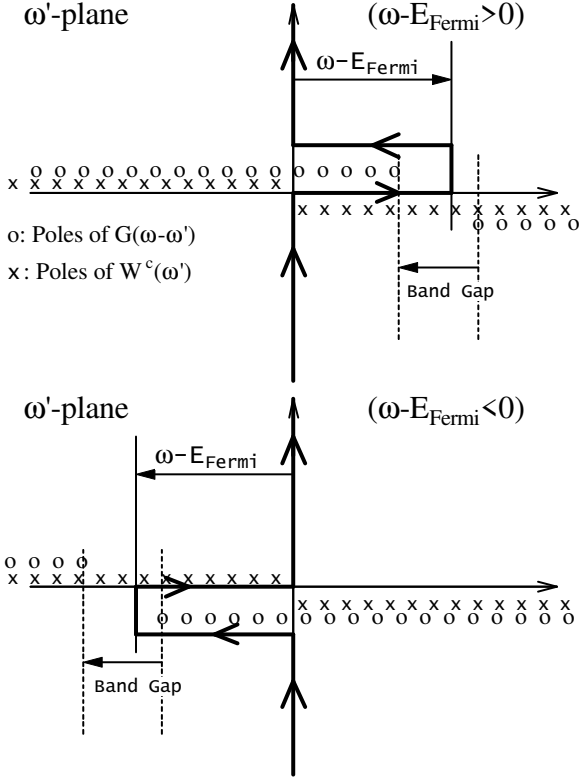} 
\end{center}
\caption[]{ Integration contour of the correlated part of the self-energy,
Eq.~(\ref{eq:sigcintI}). The original path along real axis is deformed 
without crossing poles.
$G({\bf q-k},\omega-\omega')= \sum_{n}
{1}/{(\omega-\omega'- \varepsilon_{{\bf q-k}n})}$ has poles at
$\omega'=\omega-\varepsilon_{{\bf q-k}n}$. }
\label{omegapath}
\end{figure}

\req{eq:sigc} contains the following integral
\begin{eqnarray}
X=\int_{-\infty}^{\infty} \frac{i d\omega'}{2 \pi}
W^{\rm{c}}(\omega')
\frac{1}{ -\omega'+ \we \pm i \delta},
\label{eq:sigcint}
\end{eqnarray}
where $\we=\omega-\varepsilon_{{\bf q-k}n'}$.  Here we
fix indexes $I,J,\bfk,\bfq,n'$ and omit them for simplicity.
We use a version of the imaginary-path axis 
integral method \cite{godby88,ferdibook}:
the integration path is deformed from the real-axis 
to the imaginary axis in such a way that no poles are crossed; see Fig.~\ref{omegapath}.
As a result, $X$ is written as the sum of an imaginary axis integral $X_{\rm img}$ 
and pole contributions $X_{\rm real}$.
 $X_{\rm img}$ is 
\begin{widetext}
\begin{eqnarray}
X_{\rm img}=&&\frac{-1}{\pi}\!\!\int_0^\infty \!\!\! d\omega'
   W_{\rm S}(\omega') \frac{\we}{{\omega'}^2 +\we^2} 
  +\frac{1}{\pi}\!\int_0^\infty \!\!\! d\omega'
   W_{\rm A}(\omega') \frac{\omega'}{{\omega'}^2 +\we^2},
\label{eq:sigcintI} 
\end{eqnarray}
\end{widetext}
where $W_{\rm S}(\omega')=[{W^{\rm{c}}(i\omega') \!+\! W^{\rm{c}}(-i\omega')}]/{2}$
and 
$W_{\rm A}(\omega')=[{W^{\rm{c}}(i\omega') \!-\! W^{\rm{c}}(-i\omega')}]/{2i}$.
Note that $W_{\rm A}(\omega') \ne 0$ unless time-reversal symmetry is satisfied.
$X_{\rm img}$ adds a hermitian contribution to $\Sigma^{\rm c}$.

We have to pay attention to the fact that $W_{\rm S}(\omega')$ is rather
strongly peaked around $\omega'=0$, and follow the prescription by
Aryasetiawan~\cite{ferdibook} to evaluate the first term in
Eq.~(\ref{eq:sigcintI}):
\begin{itemize}
\item[(i)] Divide $W_{\rm S}(\omega')$ into $W_{\rm S}(0) \exp(- a^2
{\omega'}^2)$ and the residual, $W_{\rm S}({\omega'}) - W_{\rm S}(0) \exp(-
a^2 {\omega'}^2)$.  $a$ is a parameter.  The first integral is performed
analytically, the residual part numerically.  We chose $a$ in
one of two ways, either to match ${d^2 W_{\rm S}(\omega')}/{d {\omega'}^2}$
at $\omega'=0$, or use $a=1.0\text{a.u.}$ (as originally done by
Aryasetiawan).  Then we find the latter is usually good enough, in the
sense that it does not impose additional burden on numerical 
integration of the residual term.

\item[(ii)] The residual term is integrated with a Gaussian quadrature in
the interval $x=[0,1]$, making the transformation $\omega' = x/(1-x)$ a.u.
(as was done first by Aryasetiawan).
Typically, a $6$ to $12$ point quadrature is sufficient to achieve convergence
less than 0.01~eV in the band gap.

\item[(iii)] On $X_{\rm img}$, we did not include smearing of the 
pole $\varepsilon_{{\bf q-k}n'}$ as explained in Sec.~\ref{sec:smear},
because it gives little effect,
although it is necessary for the evaluation on $X_{\rm real}$ as described below.
However, we add another factor to avoid a numerical problem;
we add an extra factor $1-\exp[-({{\omega'}^2+\we^2})/{2\alpha^2}]$ 
in the integrand of the numerical integration.
This avoids numerical difficulties, since ${\we}/{({\omega'}^2+\we^2)}$ can be large.
In our implementation, we simply fix $\alpha$ in (iii) to be the same 
as $\sigma$ for the smearing in \req{eq:smeargauss} and check the convergence. 
Generally speaking, $\sigma =0.001$ a.u. is satisfactory
(the differences are negligible compared to $\sigma \to 0$ results).
This procedure is not always necessary, but it makes calculations safer.
\end{itemize}

Thus the analytic part proportional to $W_{\rm S}(0)$ is 
\begin{eqnarray}
&&\frac{-W_{\rm S}(0)}{\pi}\!\!\int_0^\infty \!\!\! d\omega'
    \frac{\we \exp(-a^2 {\omega'}^2)
    \left(1-e^{  -\frac{{\omega'}^2+\we^2}{2\alpha^2} } \right)}{{\omega'}^2 +\we^2} \nonumber\\
&&=\frac{-W_{\rm S}(0)e^{a^2 \we^2}}{2} \nonumber\\
&& \times \left\{ {\rm erfc}(a \we)-{\rm erfc}\left(\we\sqrt{a^2+\frac{1}{(2\alpha^2)}} \right)\right\},
\label{eq:w0img}
\end{eqnarray}
where we use a formula $\displaystyle \int_0^\infty dx
e^{- \mu (x^2 + \beta^2)}/(x^2 + \beta^2) = \frac{\pi}{2 \beta^2} {\rm erfc}(\beta \mu)$.
Here ${\rm erfc}(...)$ is the complementary error function
(to check this formula, differentiate with respect to $\mu$ on both sides).
For small $\alpha$, we expand \req{eq:w0img} in a Taylor series
in $\alpha$, to keep numerical accuracy.

Next, let us consider $X_{\rm real}$. It has three branches
\begin{eqnarray}
X_{\rm real}=
\begin{cases}
W^{\rm{c}}(\we) & \mbox{if} \ \omega-E_{\rm F} > \we >0 \\
-W^{\rm{c}}(\we) & \mbox{if} \  0> \we >\omega-E_{\rm F} \\
0                     & \mbox{otherwise} \\
\end{cases}
\label{eq:sigcintR}
\end{eqnarray}
Poles are smeared out as discussed in Sec.~\ref{sec:kint};
$\varepsilon_{{\bf q-k}n'}$ (thus $\we$) has a Gaussian distribution.  This
means that we take $\pm wW^{\rm{c}}(\overline{\we})$ instead of
\req{eq:sigcintR}, where $w$ is the weight sum falling in the range $[0,
\omega-E_{\rm F}]$ (or $[\omega-E_{\rm F},0]$), and $\overline{\we}$ is the
mean value of $\we$ in the range.  
The QP lifetime (which originates from the
non-hermitian part of $\Sigma^{\rm c}$) comes entirely from the
imaginary part of $X_{\rm real}$. 
Thus the condition that the lifetime is infinite (no imaginary part)
at $\omega=\EF$ is assured from the path shown in Fig.~\ref{omegapath},
since $\omega=\EF$ means no contour on real axis.
The lifetime of a QP is due to its decay to another QP 
accompanied by the states included in ${\rm Im}[W^{\rm{c}}]$;
these states can be independent
electron-hole excitations, plasmon-like collective modes, local collective
charge oscillations (e.g. $d$ electrons in transition metals oscillate with
$\sim$5~eV~\cite{kotani00}), and also their hybridizations.
In the insulator case,  there is ``pair-production threshold'':
if the electron QPE is below 
$(\text{conduction band minimum}) + E_g$,
its imaginary part is zero because
it can not decay to another electron accompanied 
by an electron-hole pair (similarly for holes):
the imaginary part of the QPE for low energy electrons (holes) 
is directly interpreted as the impact ionization rate.
We will show results elsewhere; to calculate QPE lifetimes,
a numerically careful treatment is necessary--especially 
for ${\rm Im} W^{\rm{c}}$.

\subsection{Self-consistency for $\Sigma$}
\label{sec:siginterpolation}

\begin{figure}[ht]
\begin{center}
\includegraphics[scale=.7]{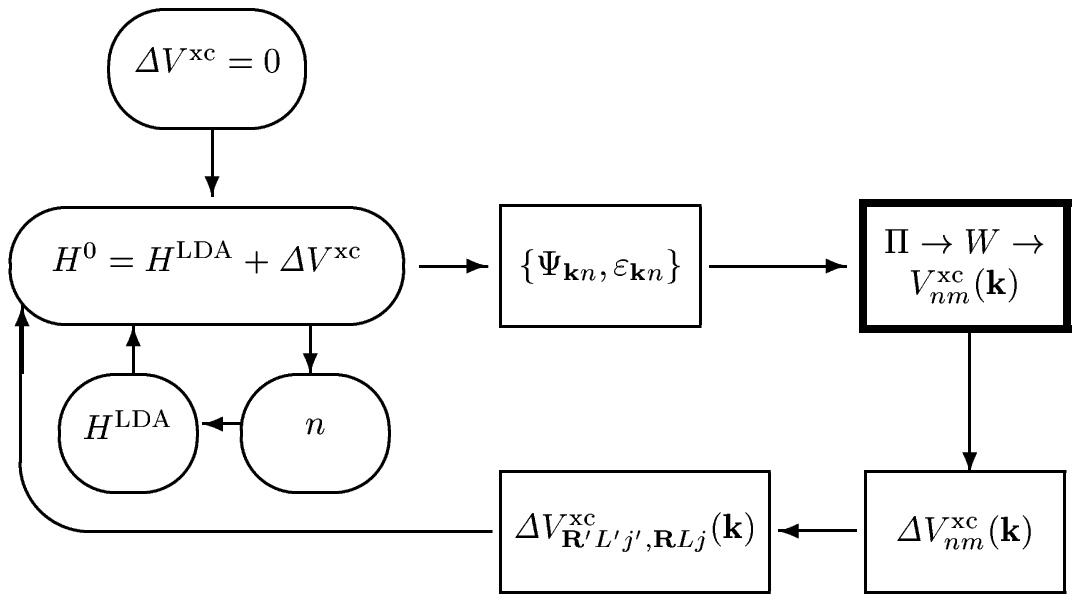} 
\end{center}
\caption[]{Self-consistency cycle.  Cycle consists of a large loop
$\HLDA+\iDelta\vxc
\to 
\Sigma({\bf{q}},\omega)\to \iDelta\vxc 
\to \HLDA+\iDelta\vxc\to\dots$.
There is an inner loop within the
\cornersize{2}\ovalbox{$\H0=\HLDA+\iDelta\vxc$} step, where the density and LDA potential are made self-consistent for a
fixed $\iDelta\vxc$.  The cycle is started by taking $\iDelta\vxc=0$, or
$\H0=\HLDA$; the first iteration is equivalent to the standard $\GLDA\WLDA$ with $Z=1$ and
including the off-diagonal parts of $\Sigma$.  The time-consuming part is
framed in bold.  Subscripts to $\Sigma$ and $\vxc$ refer to the basis it is
represented in; see text.
}
\label{fig:sccycle}
\end{figure}

The self-consistent cycle is shown in Fig.~\ref{fig:sccycle}.  It
is initiated as in a standard $GW$ calculation, by using $\HLDA$ for
$\H0$ (``zeroth iteration'').  It is a challenge to calculate the \qsgw\
$\vxc$, \req{eq:veff}, efficiently.  The most time-consuming part of a
standard $GW$ calculation is the generation of the polarization function,
$\Pi({\bf{q}},\omega)$, because only diagonal parts of $\Sigma$ are required.
However in the \scgw\ case, the generation of $\Sigma$
for \req{eq:veff} is 
typically 3 to 10 times more expensive computationally.  These two steps
together (bold frame, Fig.~\ref{fig:sccycle}) are typically 100 to 1000 times
more demanding computationally than the rest of the cycle.  Once a new
static $\vxc$ is generated, the density is made self-consistent in a
``small'' loop where the difference in the \scgw\ and LDA
exchange-correlation potentials, $\iDelta\vxc\equiv\vxc-V^{\rm xc,LDA}$, is
kept constant.  $\vxc-V^{\rm xc,LDA}$ is expected to be somewhat less
sensitive to density variations than $\vxc$ itself, so the inner loop
updates the density and Hartree potential in a way intended to minimize the
number cycles needed to generate $\Sigma$ self-consistently.  

The required number of cycles in the main loop to make $\Sigma$ self-consistent
depends on how good the LDA starting point is: for a simple semiconductor
such as Si, typically three or four iterations are enough for QP levels to
be converged to 0.01~eV.  The situation is very different for complex
compounds such as NiO; the number of cycles then depends not only on the
quality of the LDA starting point, but other complicating factors as well.
In particular, when some low-energy fluctuations (spin and orbital moments) exist,
many iterations can be required for self-consistency.
Solutions near a transition to a competing electronic state are also
difficult to converge.

$\vxc$ is calculated only at irreducible $\bfk$-points on the regular mesh,
\req{kmesh}.  However, we must evaluate $\iDelta\vxc(\bfk)$ at continuous
$\bfk$-points for a viable scheme.  In the self-consistency cycle, for
example, the offset-$\Gamma$ method requires eigenfunctions at other
${\bfk}$.  The ability to generate $\H0$ at any ${\bfk}$ is also needed to
generate continuous energy bands (essential for reliable determination of
detailed properties of the band structure such as effective masses) or
integrated quantities, e.g. DOS or EELS calculations.  Also, while it is
not essential, we typically use a finer ${\bfk}$ mesh for the step in
the cycle that generates the density and Hartree potential (oval loops in
Fig.~\ref{fig:sccycle}), than we do in the $GW$ part of the
cycle~\cite{metalnote}.

The interpolation is accomplished in the (generalized) LMTO basis by
exploiting their finite range.  Several transformations are necessary for a
practical scheme, which we now describe.  For a given 
$\iDelta\vxc$, there are three kinds of basis sets: the MTOs
$\chi^{\bfk}_{\brl{j}}$; the basis
$\Psi_{\bfk n}$ in which $\H0=\HLDA+\iDelta\vxc$ is diagonal
(the ``\scgw'' basis); and
the basis $\Psi_{\bfk \tiln}^{\rm LDA}$ in which
$\HLDA$ is diagonal (the ``LDA'' basis). 
`$\sim$' over the subscripts signifies that the function is represented
in the basis of LDA eigenfunctions.  
LDA and orbital basis sets are related by the linear transformation
Eq.~(\ref{eq:lmtopsi}).  For example $\iDelta\vxc_{nm}$ in the \scgw\ basis,
$\iDelta\vxc_{nm}({\bf k})\equiv \left< \Psi_{\bfk n} \right| \iDelta\vxc({\bf k}) \left| \Psi_{\bfk m} \right>$,
is related to
$\iDelta\vxc_{\brlp{j'},\brl{j}}({\bf k})\equiv\left< \chi^{\bfk}_{\brlp{j'}} \right|\iDelta\vxc({\bf k})\left|\chi^{\bfk}_{\brl{j{}}}\right>$
in the MTO basis by
\begin{eqnarray}
\iDelta\vxc_{nm}({\bf k})=
\sum_{\brlp{j'},\brl{j}}z^{{\bf{k}}\dag}_{\brlp{j'},n}\,\iDelta\vxc_{\brlp{j'},\brl{j}}({\bf k})\,z^{{\bf k}}_{\brl{j},m} 
 \nonumber \\
\label{eq:vxcfromvxcmto}
\end{eqnarray}
$\vxc({\bfk})$ is generated in the \scgw\ basis on the irreducible
subset of $\bfk$ points given by the mesh \req{kmesh}, which we
denote as $\bfk_{\rm mesh}$.  Generally, interpolating
$\vxc({\bfk}_{\rm mesh})$ to other ${\bfk}$ will be problematic
in this basis, 
because of ambiguities near band-crossings.  This problem
can be avoided by transformation to the MTO basis (localized in real
space).  Thus, the transformation proceeds as:
\[
\iDelta \vxc_{nm}({\bf k}_{\rm mesh}) \to 
\iDelta\vxc_{\brlp{j'},\brl{j}}({\bf k}_{\rm mesh})\to
\iDelta\vxc_{\brlp{j'},{{\bf R+T}L}{j}}.
\]
Eq.~(\ref{eq:vxcfromvxcmto}) is inverted to obtain
$\iDelta\vxc_{\brlp{j'},\brl{j}}({\bf k}_{\rm mesh})$ with the
transformation matrix $z^{{\bf{k}}}_{\brl{j},m}$.
The last step is
an inverse Bloch transform of $\iDelta\vxc_{\brlp{j'},\brl{j}}({\bf k})$ to
real space.  It is done by FFT techniques in order to exactly recover
$\iDelta\vxc_{\brlp{j'},\brl{j}}({\bf k})$ by the Bloch sum $\sum_{\bf
T}e^{i{\bf k}\cdot{\bf T}}\iDelta\vxc_{\brlp{j'},{{\bf R+T}L}{j}}$.  It is
accomplished in practice by rotating $\iDelta\vxc$ according to the crystal
symmetry to the mesh of $\bfk$-points in the full BZ, \req{kmesh}.


From $\iDelta\vxc_{\brlp{j'},{{\bf R+T}L}{j}}$, we can make
$\iDelta\vxc_{\tiln\tilm}({\bf k})$ in the LDA basis for any
$\bfk$ by Bloch sum and basis transformation.
We then approximate $\iDelta\vxc(\bfk)$
for the higher lying states by a matrix that is diagonal in the LDA basis.
The diagonal elements are taken to be linear function of the corresponding
LDA energy, $\iDelta\vxc_{\tiln\tiln}(\bfk) = a \varepsilon^{\rm
LDA}_\tiln+b$. 
Thus we use $\iDelta\vxc$ as
\begin{eqnarray*}
\begin{cases}
\iDelta\vxc_{\tiln\tilm}({\bf k}), 
& {\varepsilon^{\rm LDA}_{{\bf k}\tiln}},  
  {\varepsilon^{\rm LDA}_{{\bf k}\tilm}} < E_\xccuttwo\\
(a{\varepsilon^{\rm LDA}_{{\bf k}\tiln}}+b)\delta_{\tiln\tilm} & \mbox{otherwise}.
\end{cases}
\end{eqnarray*}
There are two reasons for this.  
First, $\iDelta\vxc_{\tiln\tilm}({\bf k})$ is well described at the mesh
points ${\bf k}_{\rm mesh}$, but often does not interpolate with enough
precision to other ${\bf k}$ for states $\tiln\tilm$ associated with very
high energy.  We believe that the unsmoothness is connected with the rather
extended range of the (generalized) LMTO basis functions in the present
implementation.  While their range is finite, the LMTOs are nevertheless
far more extended than, e.g. maximally localized Wannier functions or
screened MTOs \cite{nmto}.  Because of their rather long range, these
basis functions are rather strongly linearly dependent
(e.g. the smallest eigenvalues of the overlap matrix can be
of order $10^{-10}$).
This, when combined with small errors in
$\iDelta\vxc_{\tiln\tilm}({\bf k})$, an unphysically sharp (unsmooth)
dependence of the QPEs on ${\bf k}$ can result for points far from a mesh point.  To
pick up the normal part, it is necessary to use $E_\xccuttwo\sim{}E_{\rm
F}+2$~Ry or so.  (The largest $E_\xccuttwo$ which still results in smooth behavior
varied from case to case.  Somewhat larger $E_\xccuttwo$ can be used for
wide-band systems such as diamond; systems with heavy elements such as Bi
often require somewhat smaller $E_\xccuttwo$).

A second reason for using the ``diagonal'' approximation for high-energy
states is that it can significantly reduce the computational effort, with
minimal loss in accuracy. 

In the time-consuming generation of
$\iDelta\vxc$ (bold box in Fig.~\ref{fig:sccycle}) only a subblock of
$\iDelta\vxc_{nm}({\bf k}_{\rm mesh})$ is calculated, for states with
$\ekn,\ekm<E_\xccutone$.  This $E_\xccutone$ 
should be somehow larger than $E_\xccuttwo$ ($E_\xccuttwo$+ 1 to 2 Ry),
so as to obtain
$\iDelta\vxc_{\tiln\tilm}({\bf k})$ with smooth $\bfk$-dependence
(this eliminates nonanalytic behavior from sudden band cutoffs, which
 prohibits smooth interpolation).

$a$ and $b$ are determined by fitting them to
$\iDelta\vxc_{\tiln\tiln}(\bfk)$ which are initially generated by a
test calculation with somewhat larger $E_\xccuttwo$.  Typically the
calculated diagonal $\iDelta\vxc_{\tiln\tiln}(\bfk)$ is reasonably
linear in energy in the window $2-4$ Ry above $E_F$, and the
approximation is a reasonable one.  In any case the contribution from
these high-lying states affect QPEs in the range of interest ($E_{\rm
F}\pm{}1$~Ry) very slightly; they depend minimally on what diagonal
matrix is taken, or what $E_\xccuttwo$ is used.  Even neglecting the
diagonal part altogether ($a=0$ and $b=0$) only modestly affects
results ($<$0.1~eV change in QPEs).  We finally confirm the stability of the
final results by monitoring how QPEs depend on $E_\xccuttwo$.
Generally the dependence of QPEs on $E_\xccuttwo$ (when
$E_\xccuttwo>$2~Ry) is very weak.  Typically we take
$E_\xccuttwo\sim{}E_{\rm F}+2.5$ to $E_{\rm F}+3.5$~Ry (larger values
for, e.g. diamond).  Thus we can say the effect of the LDA used as a
reference here is essentially negligible.

$H^0$ is then given as
$H^0_{\tiln\tilm}({\bf k})={\varepsilon^{\rm LDA}
_{{\bf k}\tiln}}\delta_{\tiln\tilm}+\iDelta\vxc_{\tiln\tilm}({\bf k})$
in the LDA basis.

We use LDA as a reference not only for the above interpolation scheme, but
also to generate MTOs.  The shape of the augmented functions
$(\varphi_{Rl}$, $\dot\varphi_{Rl}$, $\varphi^z_{Rl})$ and also CORE states
are generated by the LDA potential.  However, we expect this little affects
the results; $sp$ partial waves are but little changed, and we use
local orbitals to ensure the basis is complete. CORE are sometimes tested
as ``core''.  Thus we believe that out implementation gives results which
are largely independent of the LDA.

\section{Numerical results}
\label{sec:result}
Here, we first show some convergence tests, and then show 
results for some kinds of materials to explain how \qsgw\ works.

\subsection{Convergence Test for Mixed basis}
\label{sec:result_mixed}

\begin{table*}
\begin{ruledtabular}
\begin{tabular}{lccccccccccc}
{\bf GaAs} &\GG$_{1c}$&\GG$_{15c}$& X$_{5v}$&  X$_{1c}$& X${_3c}$& L$_{3v}$&  L$_{1c}$&  L$_{3c}$\\
\multicolumn{10}{l}{(i) varying $|{\bf q+G}|^W_{\rm Max}$} \\
3.5,1.8    &   1.932  &   4.761   &  -2.973    &   2.026  &  2.452  &  -1.259  &   2.087  &   5.588  \\  
3.5,2.5            &   2.041  &   4.799  &  -2.935  &   2.168  &   2.566  &  -1.245  &   2.189  &   5.652  \\  
\underline{3.5,2.6}    &   2.046  &   4.803  &  -2.933  &   2.175  &   2.571  &  -1.244  &   2.194  &   5.657  \\
3.5,3.0            &   2.053  &   4.809  &  -2.930  &   2.186  &   2.580  &  -1.243  &   2.201  &   5.664  \\  
3.5,3.5            &   2.053  &   4.810  &  -2.929  &   2.188  &   2.581  &  -1.243  &   2.202  &   5.666  \\
\hline
\multicolumn{10}{l}{(ii) varying $|{\bf q+G}|^\Psi_{\rm Max}$} \\
2.6,2.6            &   2.038  &   4.795  &  -2.939  &   2.158  &   2.558  &  -1.247  &   2.183  &   5.647  \\ 
3.0,2.6            &   2.045  &   4.801  &  -2.935  &   2.172  &   2.569  &  -1.245  &   2.192  &   5.654  \\  
\underline{3.5,2.6}    &   2.046  &   4.803  &  -2.933  &   2.175  &   2.571  &  -1.244  &   2.194  &   5.657  \\
\hline
\multicolumn{10}{l}{(iii) varying the product basis for $W$} \\
Small       &   1.249  &   4.611  &  -3.202  &   1.702  &   2.133  &  -1.364  &   1.623  &   5.452  \\  
\underline{3.5,2.6}    &   2.046  &   4.803  &  -2.933  &   2.175  &   2.571  &  -1.244  &   2.194  &   5.657  \\
Big         &   2.057  &   4.811  &  -2.928  &   2.193  &   2.588  &  -1.243  &   2.207  &   5.669  \\  
\end{tabular}
\end{ruledtabular}
\caption{
Three tests for GaAs, showing convergence with respect to the number of the 
interstitial plane waves (IPW), and product basis(PB).
Calculations are performed for a $4\x4\x4$ ${\bf k}$-mesh, \req{kmesh}
(Results are not $k$-converged, but the differences between rows
will not change significantly.)
QPEs are in eV, relative to the top of valence.
For tests (i) and (ii), numbers in the first column are cutoffs in the 
${\bf G}$-vectors for wave functions ($|{\bf q+G}|^\Psi_{\rm Max}$) and 
Coulomb interaction ($|{\bf q+G}|^W_{\rm Max}$).
$|{\bf q+G}|=(3.5,3.0,2.6,2.5,1.8)$ a.u. correspond to 
(229,137,89,65,27) IPWs at ${\bf q}=0$.
The starting $\H0$ was generated by the \underline{3.5,2.6} case
and was not updated to perform the other tests.  
Test (iii) used the reference IPW cutoffs and varied the product basis, as
described in the text. 
}
\label{tab:gasconv}
\end{table*}

Table~\ref{tab:gasconv} shows convergence checks of the mixed basis
(product basis and IPWs) expansion of $W$, and the convergence in IPWs for
eigenfunctions.
We take GaAs, where we use 92 MTO basis set for valence:
\begin{itemize}
\item[Ga:]$4s \x 2+ 4p \x 2 + 3d \x 2+ 4f  + 5g+ 4d$(local); 39 functions 
\item[As:] $4s \x 2+ 4p \x 2 + 3d \x 2+ 4f  + 5g + 5s$(local); 35 functions
\item[Fl:]$1s  + 2p  + 3d$; 9$\times$2 functions.
\end{itemize}
``Fl'' denotes floating orbitals (MTOs with zero augmentation radius
\cite{mark06adeq}) centered at the two interstitial sites.

$\H0$ was made self-consistent for a reference case; then
one-shot calculations were performed systematically varying parameters in
the mixed basis. The result is shown in Table \ref{tab:gasconv}, 
where computational cases are specified by the two numbers 
(or labels ``Small'' and ``Big'') in the left-end column.
These are IPW cutoffs $|{\bf q+G}|^\Psi_{\rm Max}$ 
and $|{\bf q+G}|^W_{\rm Max}$ introduced in Sec.~\ref{sec:mixed}.
The reference case is denoted by \underline{3.5,2.6}
meaning $|{\bf q+G}|^\Psi_{\rm Max}=3.6$ and $|{\bf q+G}|^W_{\rm Max}=2.6$.

In the test (i) in Table \ref{tab:gasconv}, we show the convergence with
respect to $|{\bf q+G}|^W_{\rm Max}$.  $|{\bf q+G}|^W_{\rm Max}=2.6$ a.u.
is sufficient for $<$0.01eV numerical convergence.  Test (ii) is for $|{\bf
q+G}|^\Psi_{\rm Max}$; this also shows that 3.6 a.u. (even 3.0 a.u.) is
sufficient $<$0.01eV convergence.

We can characterize the PB by the number of radial functions in
each $l$ channel.  In the reference case this consists of (5,5,6,4,3) and
(6,6,6,4,2) functions for $l_{\rm p}=0,..,4$ on the Ga and As sites,
respectively ($l_{\rm p}^{\rm MAX}=4$ in this case).  The total number of
PB is then $5\x1+5\x3+6\x5+4\x7+3\x9=100$ on Ga and on As.  The
``Big'' PB of Table~\ref{tab:gasconv} used $l_{\rm p}^{\rm
MAX}=5$ with (8,8,8,6,3,2) functions for $l_{\rm p}$=0,..,5 on Ga and As (326
PB total), while the ``Small'' PB had $l_{\rm p}^{\rm
MAX}=2$ with (5,5,6) and (6,6,6) functions for $l_{\rm p}=0,1,2$ on Ga and As.  Test
(iii) shows that the reference PB is satisfactory; it is a
typical one for calculations.  Comparison with the ``Small'' case shows that
$l_{\rm p}^{\rm MAX}=2$ is poor; in our experience, $l_{\rm p}^{\rm MAX}$
must be twice larger than the maximum $l$ basis function that has
significant electron occupancy.

In addition we tested many possible kinds MTO basis sets, similar to the
tests presented for Si in Ref.~\ocite{mark06adeq}.  For GaAs, e.g.,
removing the $5g$ orbitals from Ga and As reduces the minimum band gap by
$\sim0.05$~eV; adding $4f\x1 + 5g \x1$ for floating orbitals increases it
by $\sim0.03$~eV.  In general we have found that it is difficult to attain
satisfactory numerical stability (convergence) to better than $\sim$0.1~eV
(this is a conservative estimate; we probably reach $\sim$0.05 eV or less
for a simple semiconductor such as GaAs).  This means that the numerical
accuracy is mainly controlled by the quality of the eigenfunctions input to
the $GW$ calculation, not by the cutoff parameters as tested here.

\subsection{Test of $k$ point convergence}

\begin{figure}[htbp]
\centering
\includegraphics[angle=0,scale=0.5]{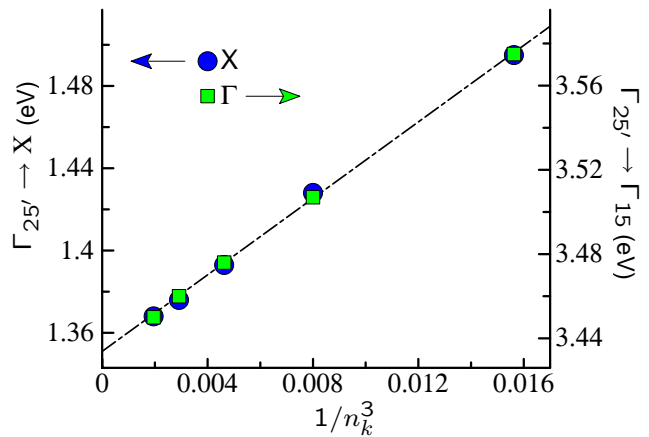}
\caption{(Color online) Convergence in the $\Gamma_{25'v}-$X$_{1c}$ (circles)
and $E_0=\Gamma_{25'v}-\Gamma_{15c}$ (squares) gaps in Si with
the density of $k$ points in the Brillouin zone, $1/n_k^3$,
calculated within the QSGW approximation.  Points correspond to
$n_k=4,5,6,7,8$.  Dot-dashed line is shown as a guide to the eye.
}
\label{fig:si_kconvergence}
\end{figure}

${\bf k}$-point convergence in \scgw calculations is somewhat harder to
attain than in LDA calculation.  
Fig.~\ref{fig:si_kconvergence} shows the
dependence of the direct and $\Gamma_{25v}\to\rm{X}_{1c}$ gaps on $n_k$,
calculated by \scgw (self-consistent results). 
Data are plotted as $1/n_k^3$, where $n_k^3=N_1N_2N_3$
is the number of mesh points in the BZ see Eq.~(\ref{kmesh}).  
Data is essentially linear in $1/n_k^3$, as it is in one-shot
calculations~\cite{mark06adeq,hamada90}.
The figure enables us to estimate the error
owing to incomplete $k$ convergence by extrapolation to $1/n_k^3\to0$.
Fig.~\ref{fig:si_kconvergence} indicates that our implementation 
is stable enough to attain the self-consistency.

\subsection{C, Si, SiC, GaAs, ZnS and ZnSe}
\label{sec:covalentsemi}

\begin{table*}
\caption{
\baselineskip 10pt
QPE relative to top of valence for various kinds of \gwa, in eV.
Calculations use $6\x6\x6$ {\bf k}-points, except rows labeled `8\x8\x8'.
$k$ convergence can be estimated by comparing `mode-A' and
`mode-A,8\x8\x8' (see also Fig.~\ref{fig:si_kconvergence}).  `1shot'
denotes standard \oneshotgw\ with $\H0=\HLDA$, \req{eq:e1shot}.  We also
show `1shotNZ' for the one-shot case with $Z$=1. `e-only' denotes
eigenvalue-only self-consistency: eigenvalues are updated but LDA
eigenfunctions are retained.
`mode-A' corresponds to \qsgw\ using \req{eq:veff}, `mode-B' to
\req{eq:veffb}.  `expt.+correction' adds to the experimental value
contributions from spin-orbit and zero-point motion (referred to as `Adj'
in Table III of Ref.~\cite{mark06adeq}).  Estimates for the latter are
taken from Table III of Ref.~\ocite{cardona05}.  Experimental data taken
from compilations in Refs. \ocite{Adachi04} and \ocite{madelung} except where noted.}
\begin{ruledtabular}
\begin{tabular}{lccccccccccc}
{\bf C}  &\GG$_{1v}$&\GG$_{15c}$&\GG$_{2'c}$& X$_{4v}$&  X$_{1c}$& L$_{3'v}$&  L$_{1c}$&  L$_{3c}$ &$E_g$\\
\colrule
LDA                & -21.32  &   5.55  &  13.55  &  -6.29  &   4.70  &  -2.79  &  9.00 & 8.39    & 4.09 \\  
1shot              & -22.24  &   7.41  &  14.89  &  -6.69  &   6.07  &  -2.99  &  10.37 & 10.34  & 5.51 \\ 
1shotNZ           & -22.60  &   7.79  &  15.26  &  -6.80  &   6.33  &  -3.04  &  10.67 & 10.76   & 5.77\\  

1shot,8\x8\x8      & -22.25  &   7.38  &  14.87  &  -6.69  &   6.04  &  -2.99  &  10.35  &  10.32  & 5.48\\
1shotNZ,8\x8\x8      & -22.62  &   7.76  &  15.23  &  -6.81  &   6.30  &  -3.05  &  10.64  &  10.73  & 5.74\\

e-only             & -23.03  &   7.92  &  15.72  &  -6.92  &   6.53  &  -3.09  &  10.96 & 10.94   & 5.94\\   
mode-A             & -23.05  &   8.01  &  15.73  &  -6.89  &   6.54  &  -3.07  &  10.96  &  10.99  & 5.97\\ 
mode-A,8\x8\x8    & -23.06  &   7.98  &  15.70  &  -6.90  &   6.52  &  -3.07  &  10.93  &  10.96  & 5.94\\ 
mode-B             & -23.05  &   8.04  &  15.79  &  -6.89  &   6.55  &  -3.07  &  10.98  &  11.00  & 5.97 \\ 
expt.              &  -23.0\ftn[1]  &   7.14\ftn[2]  &         &         &   6.08\ftn[3]     &  &      &         &  5.5\ftn[3]\\
expt.+correction   &         &         &         &         &   6.45\ftn[4]  &  &         &         &  5.87\ftn[4]  \\
\end{tabular}
\end{ruledtabular}

\begin{ruledtabular}
\begin{tabular}{lccccccccccc}
{\bf Si}  &\GG$_{1v}$&\GG$_{15c}$&\GG$_{2'c}$& X$_{4v}$&  X$_{1c}$& L$_{3'v}$&  L$_{1c}$&  L$_{3c}$ & $E_g$\\
\colrule
LDA                & -11.98   &   2.52    &   3.22    &  -2.86  &   0.59  &  -1.20  &   1.42  &   3.29   & 0.46\\
1shot              & -11.90  &   3.14  &   4.04  &  -2.96  &   1.12  &  -1.25  &   2.07  &   3.91  & 0.98\\
1shotNZ            & -11.90  &   3.34  &   4.30  &  -3.01  &   1.27  &  -1.28  &   2.26  &   4.11  & 1.13  \\
1shot,8\x8\x8      & -11.89  &   3.13  &   4.02  &  -2.96  &   1.11  &  -1.25  &   2.05  &   3.89  & 0.97\ftn[5]\\
1shotNZ,8\x8\x8    & -11.89  &   3.32  &   4.28  &  -3.01  &   1.25  &  -1.27  &   2.24  &   4.09  & 1.11\\
e-only             & -12.17  &   3.36  &   4.30  &  -3.05  &   1.28  &  -1.29  &   2.26  &   4.14  & 1.14  \\
mode-A             & -12.20  &   3.47  &   4.40  &  -3.05  &   1.38  &  -1.28  &   2.36  &   4.25  & 1.25  \\
mode-A,8\x8\x8     & -12.19  &   3.45  &   4.38  &  -3.05  &   1.37  &  -1.28  &   2.35  &   4.23  & 1.23  \\
mode-B             & -12.21  &   3.52  &   4.49  &  -3.04  &   1.42  &  -1.28  &   2.42  &   4.29  & 1.28   \\
expt.              & -12.50  &   3.35\ftn[6]&4.18&  -2.9   &   1.32  &  -1.2   &   2.18\rlap{\ftn[6]}
                                                                                         &   4.10\rlap{\ftn[6]} &  1.17  \\
expt.+correction   &         &         &         &         &         &         &         &         &  1.24\ftn[7]  \\
\end{tabular}
\end{ruledtabular}
\footnotetext[2]{Ref.~\ocite{Jimenez97}}
\footnotetext[2]{Ref.~\ocite{Logothetidis92}}
\footnotetext[3]{5.50~eV for fundamental gap on
  $\Delta$ line~\cite{Himpsel79}, adding (calculated) difference 0.58~eV from min$\to$X.}
\footnotetext[4]{EP renormalization 0.37~eV~\cite{cardona05}}
\footnotetext[5]{This number was 0.95 eV in Ref.~\ocite{mark06adeq} because of differences in computational conditions.}
\footnotetext[6]{Ellipsometry~\cite{Lautenschlager87}.  L$_c$ data assumes L$_{3'v}$=$-$1.28~eV).}
\footnotetext[7]{Correction: 0.01~eV (SO) + 0.06 (EP renormalization)~\cite{cardona05}}
\label{tab:gap1}
\end{table*}

\begin{table*}
\caption{\baselineskip 10pt
 See caption for Table \ref{tab:gap1}.}
\begin{ruledtabular}
\begin{tabular}{lccccccccccc}
{\bf SiC}          &\GG$_{1v}$&\GG$_{1c}$&\GG$_{15c}$& X$_{5v}$&  X$_{1c}$& L$_{3v}$&  L$_{1c}$&  L$_{3c}$\\
\colrule
LDA                & -15.31  &   6.25  &   7.18  &  -3.20  &   1.31  &  -1.06  &   5.38  &   7.13  \\
1shot              & -15.79  &   7.18  &   8.54  &  -3.43  &   2.16  &  -1.14  &   6.51  &   8.41  \\
1shotNZ            & -16.01  &   7.43  &   8.89  &  -3.51  &   2.36  &  -1.17  &   6.79  &   8.73  \\
1shot,8\x8\x8      & -15.79  &   7.16  &   8.53  &  -3.43  &   2.14  &  -1.14  &   6.49  &   8.39  \\
1shotNZ,8\x8\x8    & -16.01  &   7.42  &   8.87  &  -3.51  &   2.34  &  -1.17  &   6.77  &   8.71  \\
e-only             & -16.33  &   7.71  &   9.09  &  -3.56  &   2.51  &  -1.18  &   6.99  &   8.95  \\
mode-A             & -16.35  &   7.70  &   9.13  &  -3.58  &   2.53  &  -1.18  &   7.01  &   8.97  \\
mode-A,8\x8\x8     & -16.35  &   7.69  &   9.12  &  -3.57  &   2.52  &  -1.18  &   7.00  &   8.96  \\
mode-B             & -16.34  &   7.78  &   9.18  &  -3.58  &   2.58  &  -1.18  &   7.07  &   9.01  \\
expt.\ftn[1]       &         &         &         &         &   2.39  & &     &  \\
\end{tabular}
\end{ruledtabular}
\begin{ruledtabular}
\begin{tabular}{lccccccccccc}
{\bf GaAs}         & \multicolumn{1}{c}{Ga $3d$ at \GG} &\GG$_{1c}$&\GG$_{15c}$& X$_{5v}$&  X$_{1c}$& X$_{3c}$& L$_{3v}$&  L$_{1c}$&  L$_{3c}$\\
LDA                & -14.87, -14.79  &   0.34  &   3.67  &  -2.72  &   1.32  &   1.54  &  -1.16  &   0.86  &   4.58  \\
1shot              & -16.75, -16.70  &   1.44  &   4.30  &  -2.86  &   1.76  &   2.11  &  -1.22  &   1.66  &   5.14  \\
1shotNZ            & -17.64, -17.60  &   1.75  &   4.49  &  -2.91  &   1.88  &   2.26  &  -1.24  &   1.88  &   5.32  \\
1shot,8\x8\x8      & -16.75  , -16.69  &   1.41  &   4.27  &  -2.86  &   1.74  &   2.08  &  -1.23  &   1.64  &   5.12  \\
1shotNZ,8\x8\x8    & -17.65  , -17.61  &   1.70  &   4.46  &  -2.91  &   1.85  &   2.23  &  -1.25  &   1.85  &   5.29  \\
e-only             & -17.99, -17.97  &   1.69  &   4.58  &  -2.94  &   1.97  &   2.32  &  -1.26  &   1.89  &   5.44  \\
mode-A             & -18.13, -18.07  &   1.97  &   4.77  &  -2.93  &   2.15  &   2.54  &  -1.24  &   2.14  &   5.63  \\
mode-A,8\x8\x8     & -18.13, -18.07  &   1.93  &   4.74  &  -2.93  &   2.12  &   2.51  &  -1.25  &   2.11  &   5.60  \\
mode-B             & -18.10, -18.04  &   2.03  &   4.80  &  -2.94  &   2.17       &  2.56  &  -1.25  &   2.18  &   5.65  \\
expt.              &   -18.8\ftn[3]  &   1.52\ftn[5]
                                               &   4.51\ftn[5]
                                                         &  -2.80\ftn[3] &   2.11\ftn[5] &       & -1.30\ftn[3] &1.78\ftn[5] & \\
expt.+correction   &                 &   1.69\ftn[6]\\

\end{tabular}
\end{ruledtabular}
\label{tab:gap2}
\footnotetext[1]{Ref.~\ocite{LBbook}}
\footnotetext[3]{Photoemission data, Ref.~\ocite{Kraut80} (Ref.~\ocite{Ley74} for
  Ga $3d$ levels)}
\footnotetext[5]{Ellipsometry from Ref.~\ocite{Lautenschlager87b}.
                 L$_c$ and X$_c$ assume L$_{4,5v}$=$-$1.25~eV, X$_{7v}$=$-$3.01~eV (\qsgw\ results with SO)}
\footnotetext[6]{Corrections include 0.11~eV (SO) + 0.06~eV(EP)}
\end{table*}

\begin{table*}
\caption{\baselineskip 10pt
 See caption for Table \ref{tab:gap1}.}
\begin{ruledtabular}
\begin{tabular}{lccccccccccc}
{\bf $\beta$-ZnS}      &\GG$_{1v}$  & \multicolumn{1}{c}{Zn $3d$ at \GG} &\GG$_{1c}$&\GG$_{15c}$& X$_{5v}$&  X$_{1c}$& X$_{3c}$& L$_{3v}$&  L$_{1c}$&  L$_{3c}$\\
LDA                & -13.10  &  -6.44, -5.95  &   1.86  &   6.22  &  -2.23  &   3.20  &   3.89  &  -0.87  &   3.09  &   6.76  \\
1shot              & -12.92  &  -7.29,  -6.93  &   3.23  &   7.73  &  -2.34  &   4.33  &   5.36  &  -0.92  &   4.58  &   8.20  \\
1shotNZ            & -12.92  &  -7.74,  -7.42  &   3.59  &   8.15  &  -2.39  &   4.62  &   5.76  &  -0.94  &   4.97  &   8.61  \\

1shot,8\x8\x8      & -12.92  &  -7.29 ,  -6.94  &   3.21  &   7.72  &  -2.35  &   4.32  &   5.35  &  -0.92  &   4.57  &   8.19  \\
1shotNZ,8\x8\x8    & -12.93  &  -7.75 ,  -7.42  &   3.57  &   8.14  &  -2.39  &   4.61  &   5.74  &  -0.94  &   4.96  &   8.59  \\

e-only             & -13.52  &  -8.14,  -7.82  &   3.83  &   8.47  &  -2.44  &   4.91  &   6.01  &  -0.96  &   5.22  &   8.95  \\
mode-A             & -13.50  &  -8.33,  -7.97  &   4.06  &   8.69  &  -2.43  &   5.06  &   6.24  &  -0.95  &   5.47  &   9.14  \\

mode-A,8\x8\x8     & -13.50  &  -8.33,  -7.97  &   4.04  &   8.68  &  -2.43  &   5.05  &   6.23  &  -0.95  &   5.45  &   9.13  \\

mode-B             & -13.53  &  -8.31,  -7.94  &   4.13  &   8.75  &  -2.44  &   5.12  &   6.30  &  -0.95  &   5.53  &   9.20  \\
expt.              &         &  -8.7\ftn[1]    &   3.83\ftn[2] &         &      &  &         &         & \\
expt.+correction   &         &        &   3.94\ftn[3]  \\
\end{tabular}
\end{ruledtabular}
\begin{ruledtabular}
\begin{tabular}{lccccccccccc}
{\bf ZnSe}      &\GG$_{1v}$  & \multicolumn{1}{c}{Zn$3d$ at \GG} &\GG$_{1c}$&\GG$_{15c}$& X$_{5v}$&  X$_{1c}$& X$_{3c}$& L$_{3v}$&  L$_{1c}$&  L$_{3c}$\\
LDA                & -13.32  &  -6.64,  -6.27  &   1.05  &   5.71  &  -2.21  &   2.82  &   3.32  &  -0.88  &   2.36  &   6.30  \\
1shot              & -13.15  &  -7.66,  -7.39  &   2.31  &   6.82  &  -2.35  &   3.62  &   4.41  &  -0.94  &   3.57  &   7.32  \\
1shotNZ            & -13.09  &  -8.17,  -7.94  &   2.63  &   7.14  &  -2.40  &   3.82  &   4.70  &  -0.96  &   3.89  &   7.62  \\
1shot,  8\x8\x8    & -13.16  &  -7.66,  -7.39  &  2.28  &   6.80  &  -2.35  &   3.60  &   4.40  &  -0.94  &   3.56  &   7.31  \\
1shotNZ,8\x8\x8    & -13.10  &  -8.18, -7.94  &   2.59  &   7.12  &  -2.40  &   3.80  &   4.68  &  -0.96  &   3.86  &   7.60  \\
e-only             & -13.61  &  -8.52,  -8.30  &   2.79  &   7.41  &  -2.43  &   4.08  &   4.92  &  -0.97  &   4.08  &   7.93  \\
mode-A             & -13.64  &  -8.63,  -8.36  &   3.11  &   7.69  &  -2.42  &   4.32  &   5.20  &  -0.96  &   4.39  &   8.19  \\
mode-A,8\x8\x8     & -13.65  &  -8.63,  -8.36  &  3.08  &   7.68  &  -2.42  &   4.30  &   5.19  &  -0.96  &   4.38  &   8.17  \\
mode-B             & -13.65  &  -8.65,  -8.38  &   3.16  &   7.72  &  -2.42     &   4.34  &   5.23  &  -0.96  &   4.43  &   8.21  \\
expt.              &         &  -9.0\ftn[4]&   2.82\ftn[5] &         &  -2.1\ftn[4] &  4.06\ftn[6] &     &  -1.2\ftn[4]   & 3.96\ftn[6]\\
expt.+correction   &         &                 &   3.00\ftn[7] \\
\end{tabular}
\end{ruledtabular}
\label{tab:gap3}
\footnotetext[1]{Ref.~\ocite{Zhou97}}
\footnotetext[2]{Ref.~\ocite{Mang96}}
\footnotetext[3]{Corrections include 0.02~eV (SO) + 0.09~eV(EP)~\cite{cardona05}}
\footnotetext[4]{Ref.~\ocite{vesely72}}
\footnotetext[5]{Ref.~\cite{Mang96}}
\footnotetext[6]{Reflectivity from Ref.~\ocite{Markowski94}.
                 L$_c$ and X$_c$ assume L$_{4,5v}$=-0.95~eV, X$_{7v}$=-2.47~eV (\qsgw\ results with SO)}
\footnotetext[7]{Corrections include 0.13~eV (SO) + 0.05~eV(EP)~\cite{cardona05}}
\end{table*}

Table \ref{tab:gap1}-\ref{tab:gap3} show QPE within various kinds of \gwa,
for C and Si, and for SiC, GaAs, ZnS and ZnSe in the zincblende structure.
`1shot' denotes the standard \oneshotgw\ from the LDA including the $Z$
factor, \req{eq:e1shot}; we also show
`1shotNZ' starting from the LDA, but using \req{eq:e1shot} with $Z$=1.
As noted by ourselves and
others~\cite{kotani02,Usuda04,Fleszar05,friedrich06,mark06adeq}, standard `1shot'
generally underestimates band gaps relative to experiment; for example, the
difference between `1shot,8\x8\x8' and LDA is $0.97-0.46$ eV for bandgap of Si.  This
difference is only $\sim$60\% of the correction $1.24-0.46$ eV needed for the
calculation to be exact.  The systematic tendency to underestimate gaps 
in `1shot' is also confirmed by a recent FP-LAPW $GW$ calculation~\cite{friedrich06} 
with huge basis sets.  `1shotNZ' shows better agreement
with experiment; a justification for using $Z$=1 was presented in
Ref.~\ocite{mark06adeq}.  
However, this good agreement is somewhat fortuitous
because of cancellation between two contributions
to $W$:  there is one contribution from the LDA bandgap underestimate,
which overestimates $\Pi$ and underestimates $W$, and another
from omission of excitonic effects which overestimates $W$\cite{mark06adeq}.
In other words, `1shot' agreement with experiment would worsen
if $W$ were to include excitonic effects:
the screening would be enhanced and result in smaller band gaps.
`e-only' denotes self-consistency in
eigenvalues, while retaining LDA eigenfunctions (and charge density).  
`mode-A' is \qsgw\ with
\req{eq:veff}, `mode-B' with \req{eq:veffb}.  `mode-A,8\x8\x8' is
$k$-converged to $\sim$0.01~eV; see Fig.~\ref{fig:si_kconvergence}.

Tables \ref{tab:gap1}-\ref{tab:gap3} shows that `e-only' results are closer
to `1shotNZ' than `1shot'.  We present results for a few materials here,
but find that it appears to be true for a rather wide range of materials.
We have already argued this point in Ref.~\ocite{mark06qsgw}.  Mahan
compares the two in the context of the Fr\"olich Hamiltonian
\cite{mahan90}.  The Rayleigh-Schr\"odinger perturbation theory corresponds
to `1shotNZ', which he argues is preferable to the Brillouin-Wigner
perturbation which corresponds to `1shot'.

`e-only' and `mode-A' do not differ greatly in these semiconductors,
because the LDA eigenfunctions are similar to the \scgw\ eigenfunctions
(this is evidenced by minimal differences in the LDA and \scgw\ charge
densities).  `mode-A' band gaps are systematically larger than experiments,
which we attribute to the omission of excitonic contributions to the
dielectric function.  
'mode-A' and `mode-B' show only slight differences, which is indicator of
the ambiguity in the \qsgw\ scheme due to the different possible choices
for the off-diagonal contributions to $\vxc$. This shows the ambiguity
is not so serious a problem in practice.

Fundamental gaps are precisely known; other high-lying indirect gaps are
well known only in a few materials.  Much better studied are direct
transitions from measurements seen as critical points in the dielectric
function or reflectivity measurements.  We present some experimental data
for states at $\Gamma$, and also X and L.
Agreement is generally very systematic: $\Gamma${-}$\Gamma$ and
$\Gamma${-}L transitions are usually overestimated by $\sim$0.2~eV, and
$\Gamma${-}X transitions a by little less.  We can expect valence band
dispersions to be quite reliable in \qsgw; they are already reasonably good
at the LDA level.  Except for a handful of cases (GaAs, Si, Ge, and certain
levels in other semiconductors) there remains a significant uncertainty in
the QP levels other than the minimum gap.  In ZnSe, for example, the L
point was inferred to be 4.12~eV from a fast carrier dynamics
measurement~\cite{Dougherty97}, while optical data show $\sim$0.2~eV
variations in $E_1$=L$_{1c}$-L$_{3'v}$; compare, for example
Refs.\ocite{Adachi91} and \ocite{Markowski94}.  Most of the photoemission
data (used for occupied states in Tables \ref{tab:gap1}-\ref{tab:gap3}) is
also only reliable to a precision of a few tenths of an eV.

Finally, when comparing to experiment, we must add corrections for
spin-orbit coupling and electron-phonon (EP) renormalization of the gap
owing to zero-point motion of the nuclei.  Neither are included in the QP
calculations, and both would reduce the \qsgw\ gaps.  EP renormalizations
have been tabulated for a wide range of semiconductors in
Ref.~\ocite{cardona05}.  In some cases there appears to be some
experimental uncertainty in how large it is.  In Si for example, the EP
renormalization was measured to be $\sim$0.2~eV in Ref.~\cite{Hidenori00},
while the studies of Cardona and coworkers put it much
smaller~\cite{cardona05}.

\subsection{ZnO and Cu$_2$O}

\begin{table}
\caption{
\baselineskip 12pt
Direct gap (eV) at $\Gamma$ for ZnO and Cu$_2$O for kinds of \gwa.
For Cu$_2$O, E$_g$ and E$_0$ are the first and second minimum gap at $\Gamma$.}
\begin{tabular}{r|r|r|r|r|r|r|r|r|r|r|r} 
\colrule
Material
& Expt.
        &  LDA  &  1shot  &1shotNZ & 1shotNZ        & e-only & mode-A \\
& +corr
        &       &         &        &        (off-d) &        &        \\
\colrule
ZnO     &  3.60\ftn[1]            &  0.71 &  2.46  &2.88 & 3.00     & 3.64     & 3.87  \\
\colrule
Cu$_2$O   E$_g$ &  2.20\ftn[1]    &  0.53 &  1.51  &1.99 & 1.95     & 1.98   & 2.36  \\
          E$_0$ &  2.58\ftn[1]    &  1.29 &  1.88  &1.97 & 1.93     & 2.32   & 2.81  \\
\colrule
\end{tabular}
\footnotetext[1]{Values computed from 3.44+0.164~eV, 2.17+0.033~eV, and
2.55+0.033~eV, where 0.164eV and 0.033eV are zero-point contributions; See
Table III in Ref.~\ocite{cardona05}.  The 1shot value 2.46~eV for ZnO is
slightly different from a prior calculation (2.44~eV in
Ref.~\ocite{usuda02}) because of more precise computational conditions.
`1shotNZ(off-d)' denotes a 1shot calculation including the off-diagonal
elements.}
\label{tab:znogap}
\end{table}

\begin{table}
\caption{
\baselineskip 12pt
Optical dielectric constant $\epsilon_\infty$ calculated in the RPA~\cite{epsnote}
from `mode-A' $\H0$.  Calculations were checked for $k$-convergence; data
shown used 3888 $k$-points for ZnO and 1444 points for the others.
``With LFC'' means including the local field correction (See e.g.~\cite{arnaud01}).}
\begin{tabular}{l|*3{@{\hspace*{0.5em}}l}}
\vbox{\vskip 2pt}
           &  no LFC & with LFC & Expt.\\
\colrule
   ZnO($\bfk$/\!/C-axis)&  3.2  &  3.0 & 3.75-4.0\ftn[1]  \\
   Cu$_2$O &     5.9  &  5.5  & 6.46\ftn[2]  \\
   MnO     &     3.9  &  3.8  & 4.95\ftn[3] \\
   NiO     &     4.4  &  4.3  & 5.43\ftn[4],5.7\ftn[3]\\
\end{tabular}
\footnotetext[1]{Ref.~\ocite{Adachi04}}
\footnotetext[2]{Ref.~\ocite{hodby76}}
\footnotetext[3]{Ref.~\ocite{mochizuki89,plendl69}}
\footnotetext[4]{Ref.~\ocite{powell70}}
\label{tab:eps}
\end{table}

\begin{figure}[htbp]
\centering
\includegraphics[angle=0,scale=0.4]{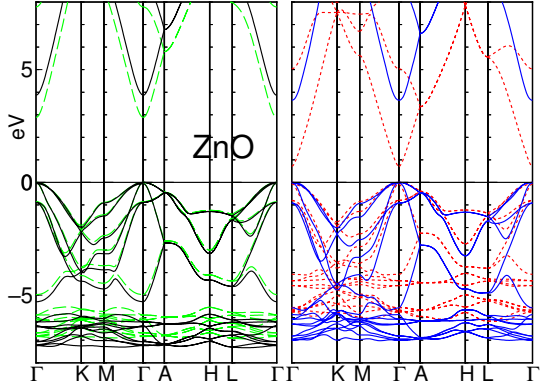}
\caption{(Color online) Energy bands in wurzite ZnO.
Solid(Black) in left panel:\qsgw(`mode-A') ;  \ 
Broken(Green) in left:1shotNZ(off-d) ; \ 
Solid(Blue) in right panel:`e-only' ; \ 
Dotted(Red) in right panel:LDA.
}
\label{fig:zno_band}
\end{figure}

\begin{figure}[htbp]
\centering
\includegraphics[angle=0,scale=0.5]{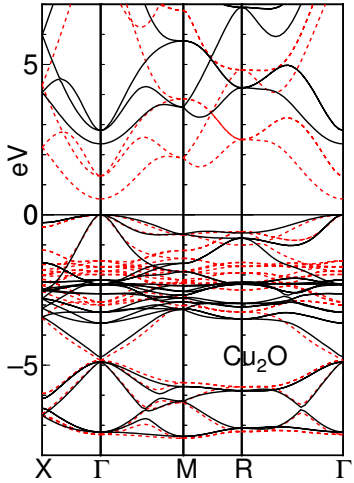}
\caption{(Color online) Energy bands of Cu$_2$O.  
Dotted(Red):LDA ; \ Solid(Black):\qsgw(`mode-A').
}
\label{fig:cu2o_band}
\end{figure}

Data for the direct gap at $\Gamma$ for wurzite ZnO, and for Cu$_2$O
(cuprite structure) are given in Table \ref{tab:znogap}.  In these
materials, `1shotNZ' (and more so `1shot') results are rather poor, as were
ZnS and ZnSe (Table~\ref{tab:gap3}).  Corresponding energy bands are shown
in Figs.~\ref{fig:zno_band} and \ref{fig:cu2o_band}.  We used 144 and 64
$\bfk$ points in the 1st BZ to make $\H0$ for ZnO and Cu$_2$O, respectively.
`1shotNZ(off-d)' denotes a 1shot calculation, including the off-diagonal
elements of $\iDelta\vxc$ computed in `mode-A'; i.e. bands were generated
by $\HLDA+\iDelta\vxc$ without self-consistency.  They differ little from
standard `1shotNZ' results in semiconductors, as we showed for variety of
materials in Ref.~\ocite{mark06adeq}.  Calculated dielectric constants
$\epsilon_\infty$ are shown in Table \ref{tab:eps}.
Note its systematic underestimation. 
Data ``With LFC'' is the better calculation, as it 
includes the so-called local field correction (See e.g.~\cite{arnaud01}).

\subsubsection{\bf ZnO}

In contrast to cases in Sec.~\ref{sec:covalentsemi}
`e-only' and `1shotNZ' now show sizable differences.  
This is because the LDA gap is much too small, 
and $W$ is significantly overestimated. 
The discrepancy is large enough that `1shotNZ,' which
approximately corresponds to `e-only' but neglecting changes in $W^{\rm
LDA}$, is no longer a reasonable approximation.  On the other band,
Table~\ref{tab:znogap} shows that the `e-only' and `mode-A' are not so
different (3.64~eV compared to 3.87~eV).  This difference measures the
contribution of the off-diagonal part to band gap; it is similar to ZnS.
This modest difference suggests that the LDA eigenfunctions are still
reasonable.  However, there still remains a difficulty 
in disentangling $d$ valence bands from the others 
in the `e-only' case: topological connections in band
dispersions can not be changed from the LDA topology, as we discussed in
Ref.~\ocite{mark06adeq}.

\begin{figure}[htbp]
\centering
\includegraphics[angle=0,scale=0.4]{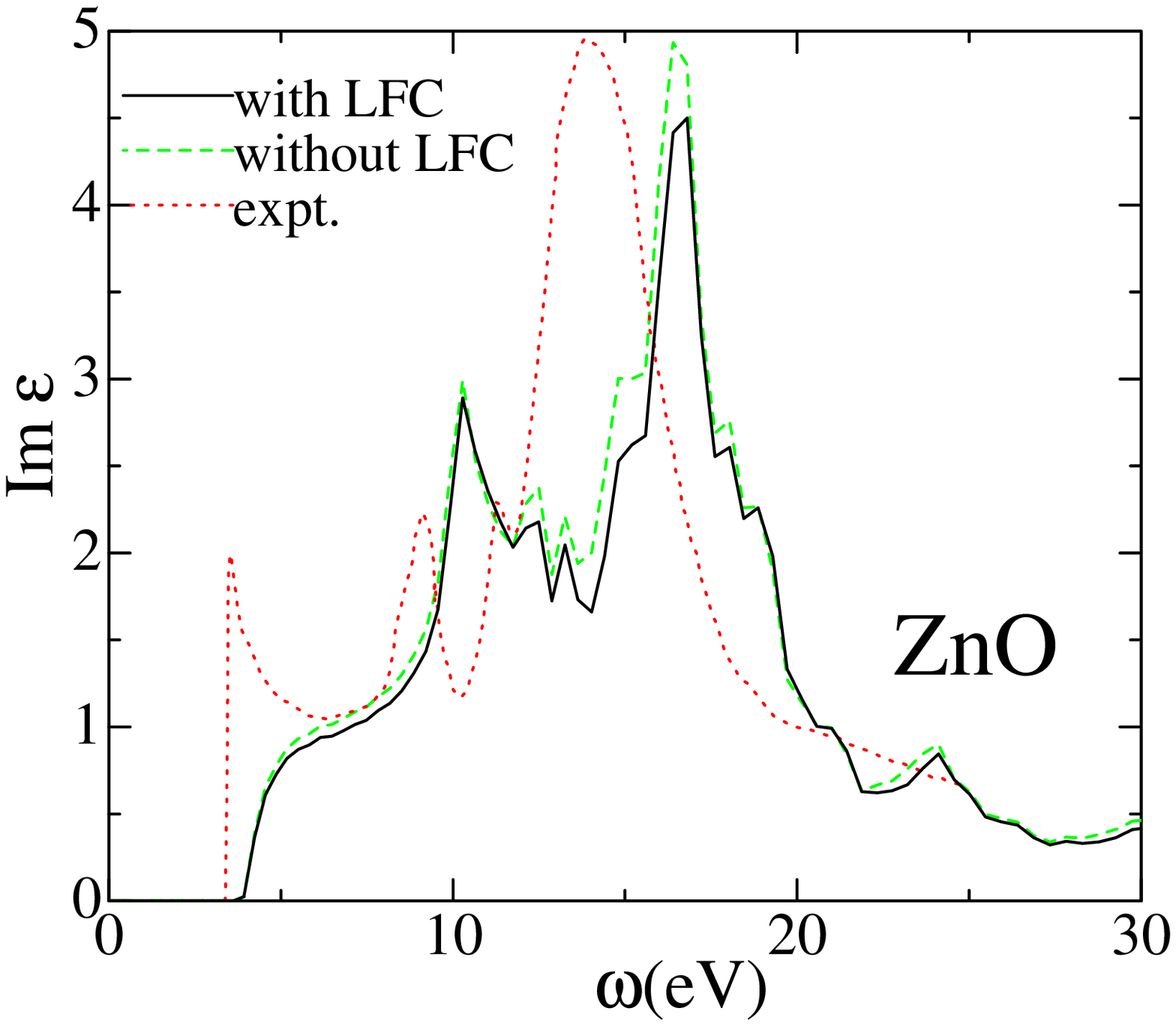}
\caption{(Color online)
Imaginary part of dielectric function for ZnO.
Local field corrections (LFC) only slightly affect the result.
}
\label{fig:zno_eps}
\end{figure}

The imaginary part of the dielectric function,
${\rm{Im}}\,\epsilon({\bf{q}}$=${0},\omega)$, is calculated from the
`mode-A' potential and compared to the experimental function in
Fig.~\ref{fig:zno_eps}.  There is some discrepancy with experiment.  Arnaud
and Alouani~\cite{arnaud01} calculated the excitonic contribution to ${\rm
Im}\,\epsilon(\omega)$ with the Bethe-Salpeter equation for several
semiconductors. Generally speaking, such a contribution can shift peaks
in ${\rm{Im}}\,\epsilon({\bf{q}}$=${0},\omega)$ to lower energy and create new
peaks just around the band edge.  It is in fact just what is
needed to correct the discrepancy with experiments as seen in
Fig.~\ref{fig:zno_eps}. 

\begin{figure}[htbp]
\centering
\includegraphics[angle=0,scale=0.4]{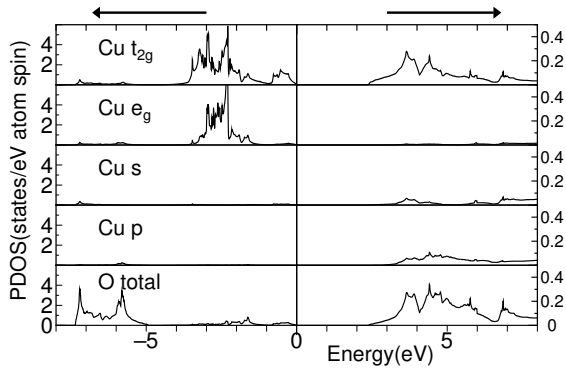}
\caption{
Partial DOS (QP-DOS) computed from the \qsgw\ $\H0$ for
Cu$_2$O.  Valence band maximum at zero.  Right-hand scale is for unoccupied
states and is 10$\x$ the left-hand scale, which applies to occupied states.
}
\label{fig:cu2o_dos}
\end{figure}

\begin{figure}[htbp]
\centering
\includegraphics[angle=0,scale=0.4]{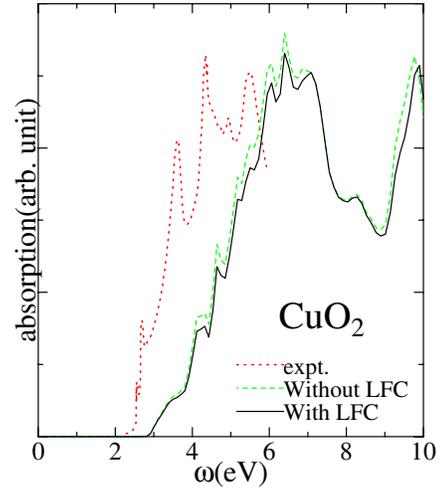}
\caption{(Color online)
Absorption coefficient for Cu$_2$O.  Experiment is taken from
Ref.~\ocite{bahms66}.  }
\label{fig:cu2o_absorb}
\end{figure}

\subsubsection{\bf Cu$_2$O}
We have to distinguish between the QP density of states (QP-DOS), 
which is calculated from $\H0$ (the poles of $G^0$), and the ``spectrum''
density-of-states, which is calculated from the poles of $G$. 
The QP-DOS is the important quantity needed to describe 
the fundamental excitations in materials.
QP-DOS in `mode-A' are shown in Fig.~\ref{fig:cu2o_dos}.
The absorption coefficient in RPA from $\H0$ is shown in
Fig.~\ref{fig:cu2o_absorb}, 
as well as the energy bands in Fig.~\ref{fig:cu2o_band}.  
This calculation includes Cu3$p$ and Cu4$d$ 
as VAL states using local orbitals.

As shown in Table~\ref{tab:znogap}, the discrepancy between ``e-only'' and
``mode-A'' fundamental gap is $\sim$0.4~eV, rather significant and larger
than cases considered previously.  This reflects an increased discrepancy
between the LDA and \qsgw\ eigenfunctions.  The `mode-A' energy bands can
be compared with results by Bruneval et al.~\cite{bruneval06cu}, who also
performed \qsgw\ calculations within a pseudopotential framework.  The
lowest and second gaps we obtain, $E_g$=2.36~eV and $E_0$=2.81~eV, are
somewhat larger than their values (1.97~eV and 2.27~eV).  Further, the
difference between the peaks just below \efermi\ and the main $3d$ peak is
1.90 eV, which is slightly larger than what Bruneval et al. obtain
(1.64~eV).  This is the D1-F1 difference in Ref.~\ocite{shen90}, measured
to be 1.94~eV.
The absorption coefficient shows essentially the same kinds of discrepancy
with experiment as we saw for Im$\,\epsilon(\omega)$ in ZnO.  Bruneval et
al. calculated the excitonic contributions for Cu$_2$O in a Bethe-Salpeter
framework~\cite{bruneval06cu}, and showed that they account for most of the
error in the RPA dielectric function as computed by \qsgw.

\subsection{NiO and MnO}

\begin{figure}[htbp]
\centering
\includegraphics[angle=0,scale=0.4]{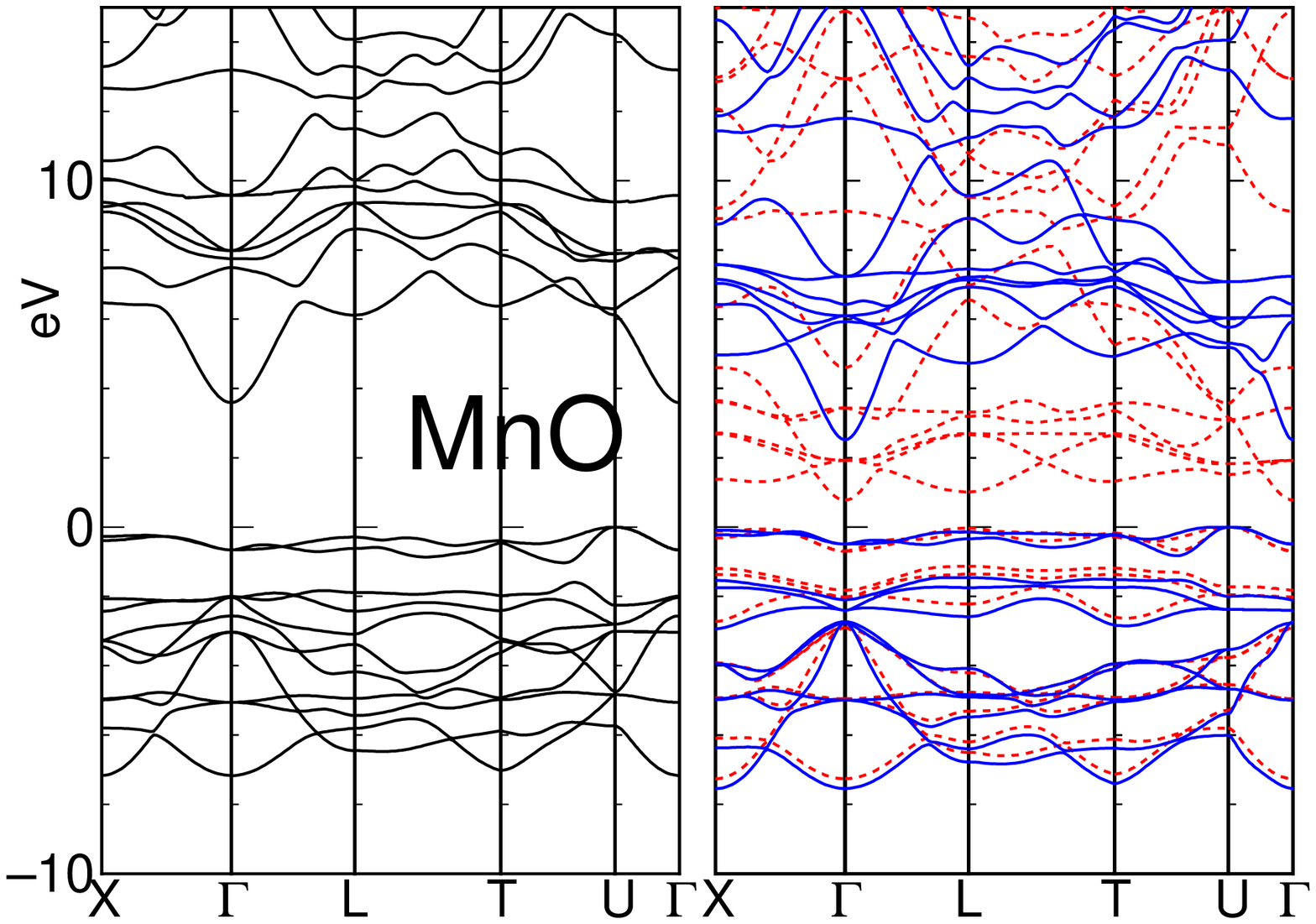}
\includegraphics[angle=0,scale=0.4]{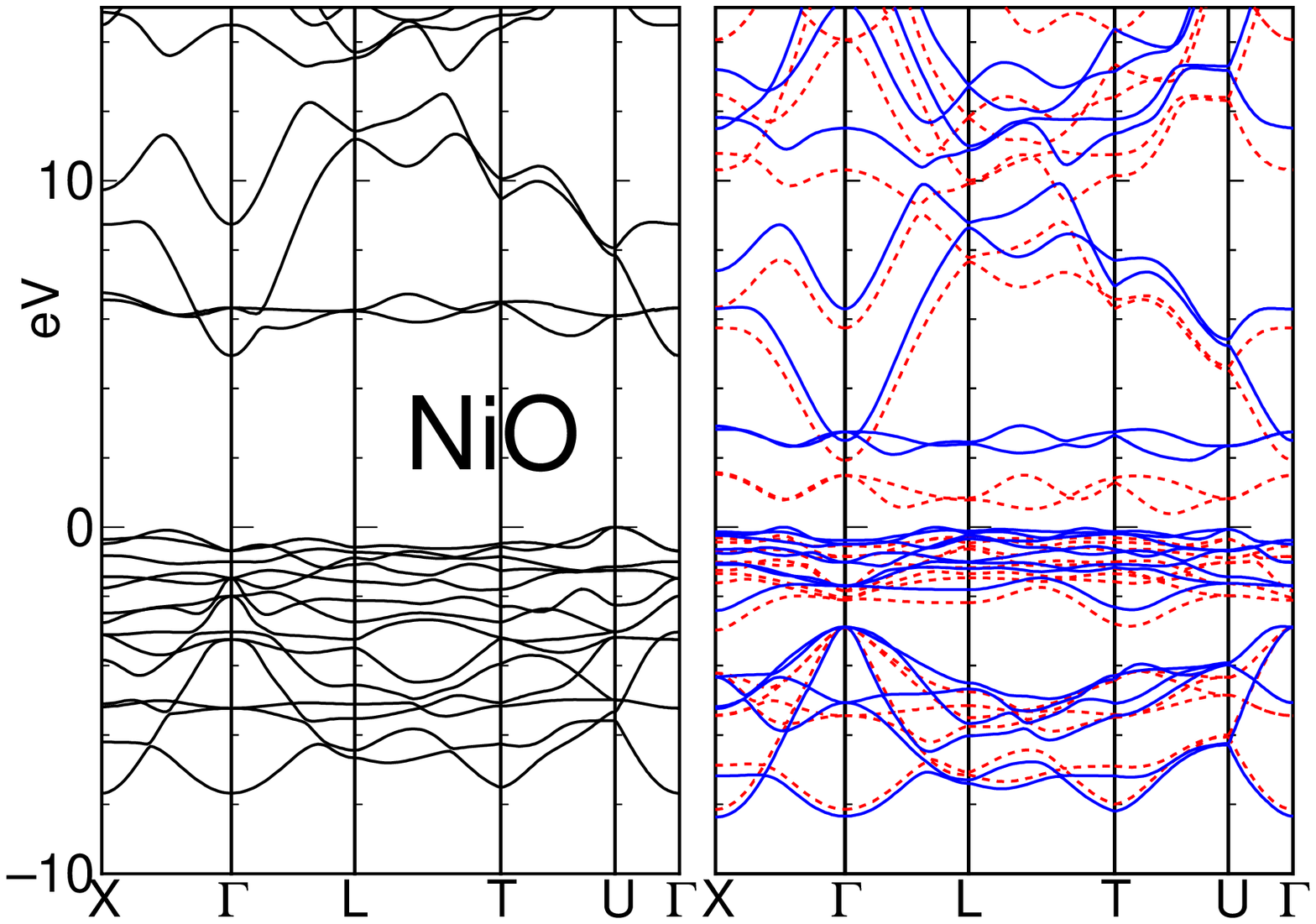}
\caption{(Color online)
Energy bands for NiO and MnO.  
Solid(Black) in left panel: \scgw(`mode-A') ;\ 
Solid(Blue) in right panel:`e-only' ;\ 
Dotted(Red) in right panel: LDA.
}
\label{fig:mno_band}
\end{figure}

\begin{figure}[htbp]
\centering
\includegraphics[angle=0,scale=0.5]{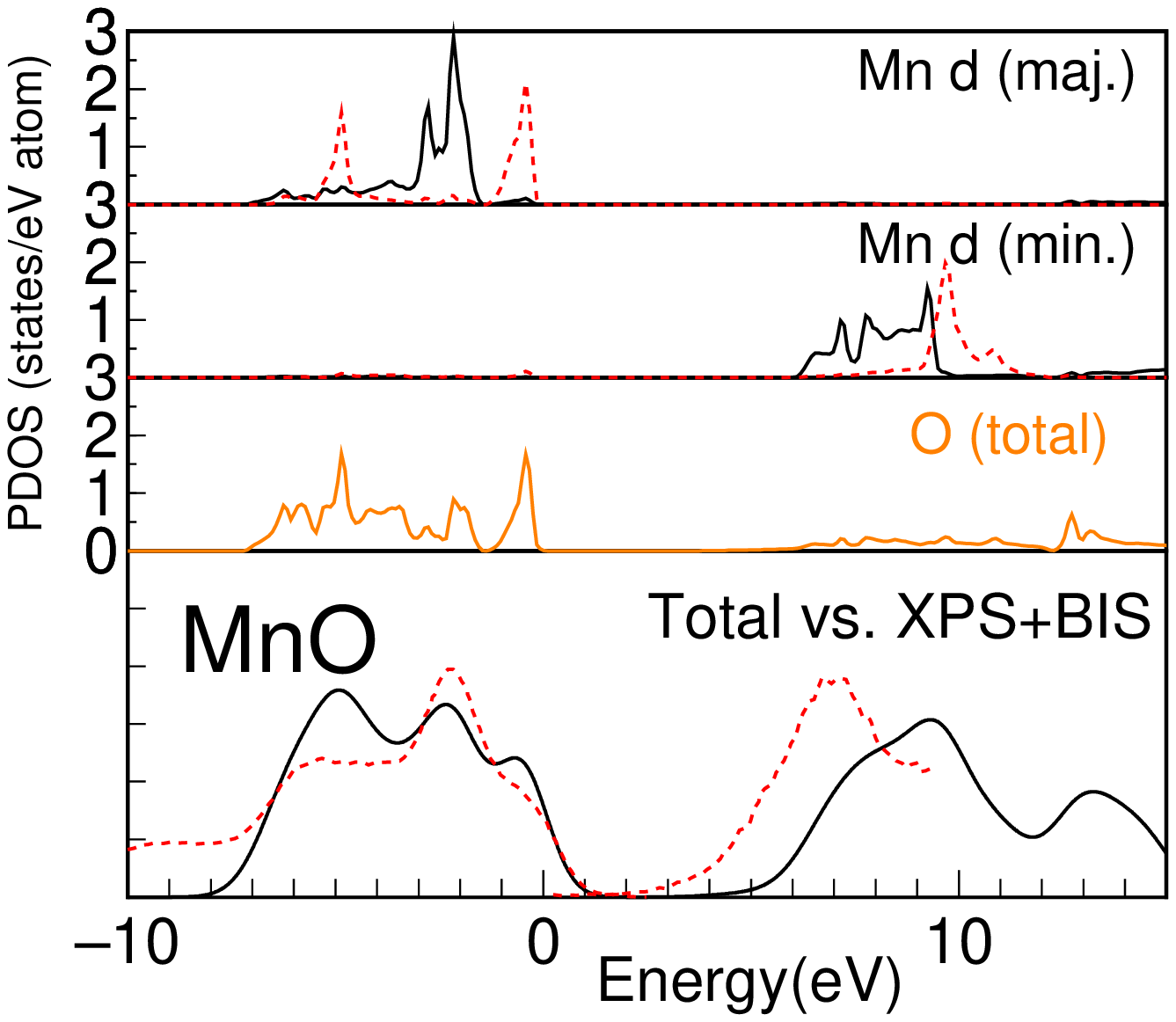}
\includegraphics[angle=0,scale=0.5]{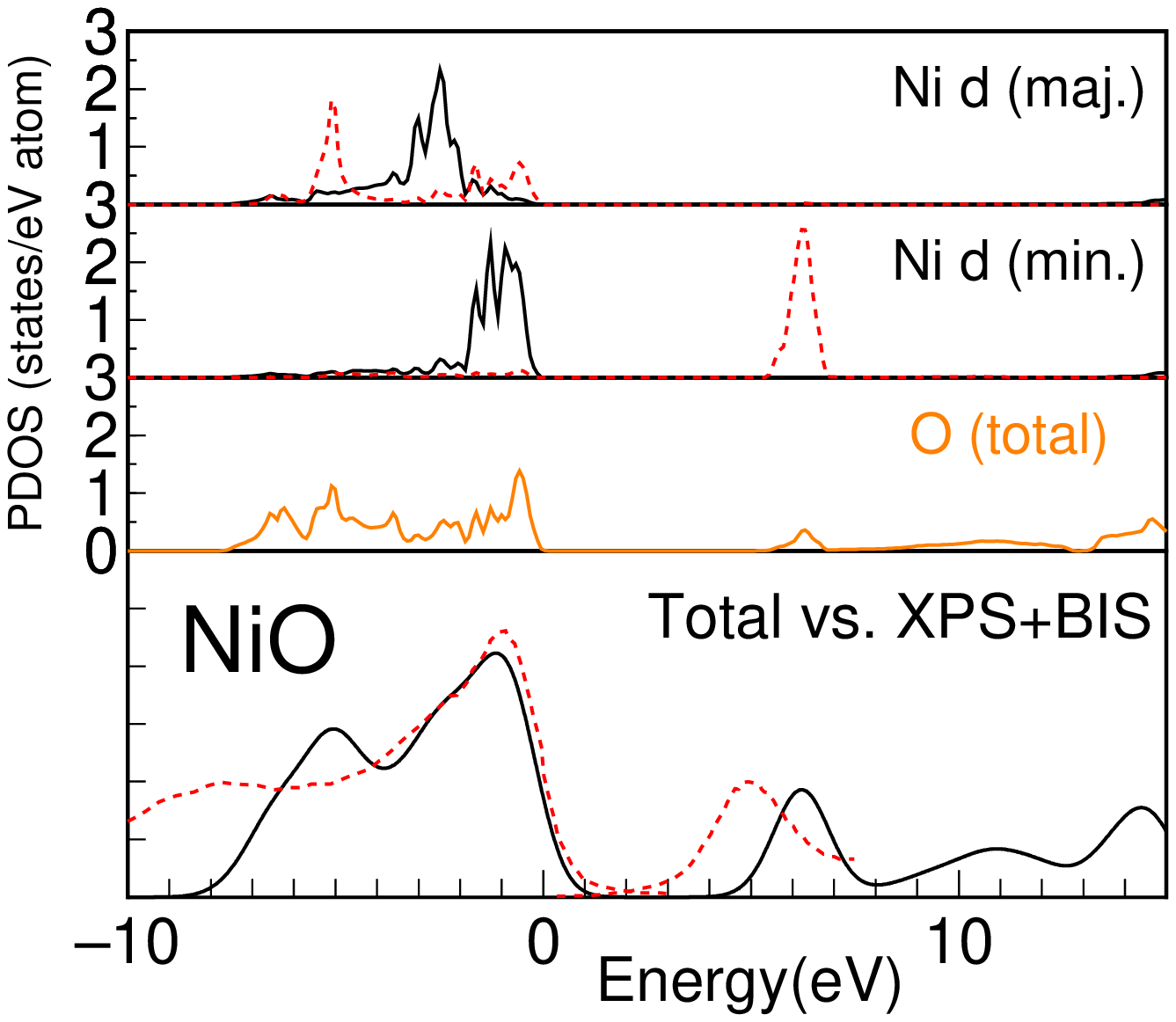}
\caption{(Color online)
QP-DOS(`mode-A') for NiO and MnO for each atomic site.
Contributions are decomposed majority and minority in Mn(Ni) site
in top two panels. Further, it is decomposed into $e_g$ and $t_{2g}$.
Dotted(Red): $e_g$; \ Solid(Black):$t_{2g}$.
O(total) is the QP-DOS project onto an O site 
(sum of majority and minority).
Dots(Red) in the bottom panels are taken from XPS+BIS 
experiments~\cite{vanelp91,vanelp92},
and are compared with total QP-DOS broadened with a 0.6~eV Gaussian.
}
\label{fig:mno_dos}
\end{figure}

We described these compounds already in Ref.~\ocite{faleev04}; here we
present some additional analysis. We assume antiferro-II
ordering~\cite{terakura84} and time-reversal symmetry (thus no orbital
moments), with 64 $\bfk$-points in the 1st BZ.  In Fig.~\ref{fig:mno_band}, we
show \scgw\ `mode-A' energy bands, comparing them to `e-only' and LDA in
the right panel.  The problem with `e-only' bands is now very apparent: the
bandgap is much too small and the conduction band dispersions are
qualitatively wrong (minimum gap falls at the wrong point for NiO).
Further, the valence bands (especially, the relative position of O 2$p$ and
TM $3d$ bands) changes little relative to the LDA.  `e-only' can not
change the LDA's spin moment since eigenfunctions do not change; they
are significantly underestimated.  The QP-DOS are shown in
Fig.~\ref{fig:mno_dos}. As we detail below, the self-consistency
is essential for these materials as was found 
by Aryasetiawan and Gunnarsson \cite{aryasetiawan95nio}.

\begin{figure}[htbp]
\centering
\includegraphics[angle=0,scale=0.4]{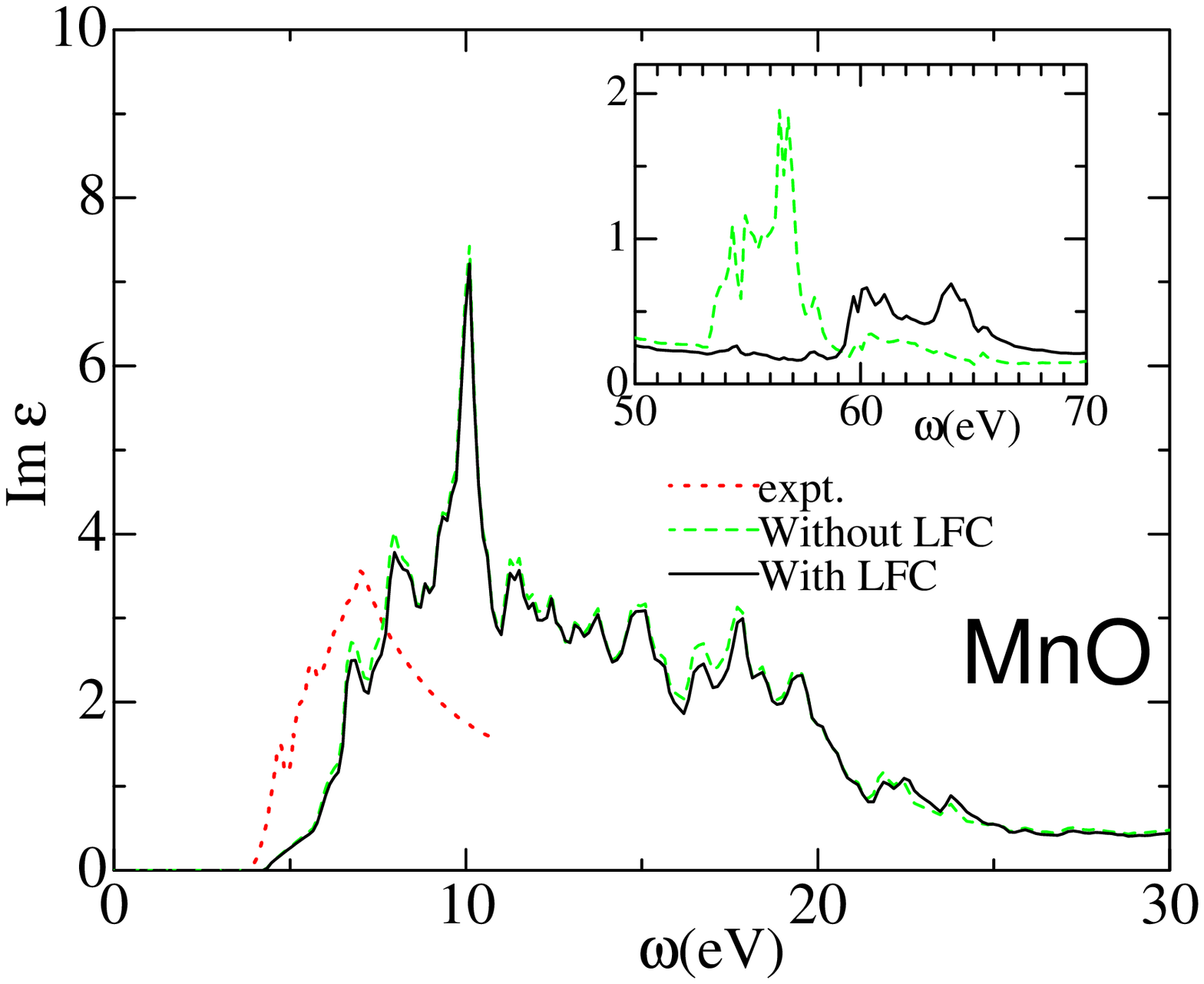}
\includegraphics[angle=0,scale=0.4]{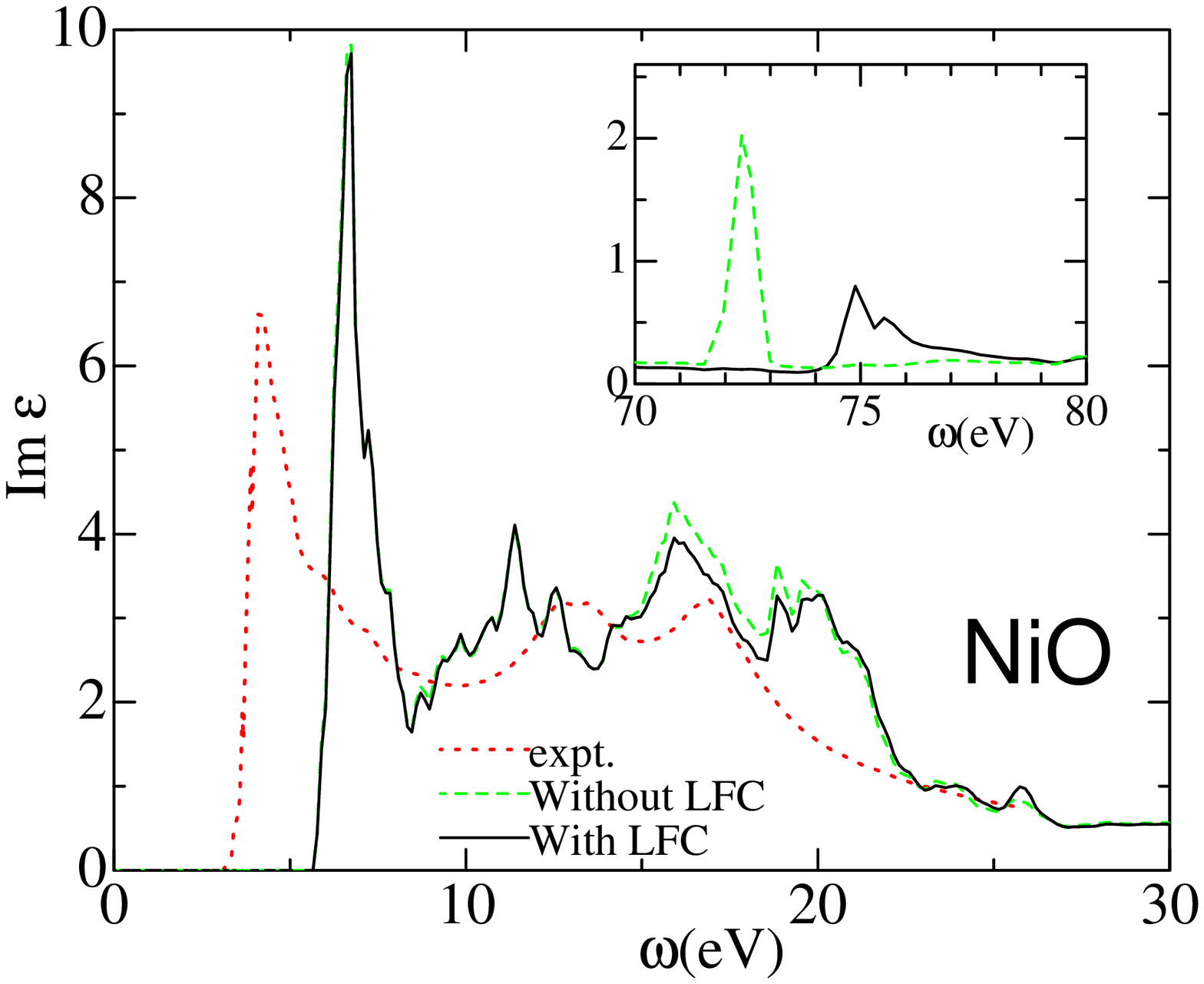}
\caption{(Color online) Imaginary part of the dielectric function in MnO and NiO.  Effects
of local fields (LFs) are significant only in the deep energy region
related to the $3p$ core (inset).  LF effects contribute 
mainly to such local atomic-like excitations; they
should be interpreted in terms of the Lorenz
field in the Clausius-Mossotti relation. 
Local fields reduce the peaks of localized excitations,
and moves them to higher energy
(this is already seen in rutile TiO$_2$ \cite{vast02}). 
Experimental data is taken from Ref.~\ocite{Messick72} for MnO, and
Ref.~\ocite{powell70} for NiO.}
\label{fig:mno_eps}
\end{figure}

\subsubsection{\bf MnO}

The Mn $e_g$ and O $2p$ QP-DOS show common peaks below \efermi.  This is
because of the strong $dd_{\sigma}$ coupling between $e_g$, mediated through
the O $2p$ valence bands~\cite{terakura84}.  The $e_g$ components are
separated mainly into two peaks which have comparable weight, in contradiction to
the LDA case where the deeper peak has a very small weight.  This is
because $e_g$ levels are pushed down relative to the LDA.  The bottom panel
compares XPS (occupied states) and BIS (unoccupied states) experiments with
the total QP-DOS, broadened with 0.6~eV Gaussian.  There is good agreement
with the XPS part for the $e_g$ peak just below \efermi, for the $t_{2g}$
peak, and for the valence band width.  (Based on a model analysis by
Takahashi and Igarashi \cite{takahashib}, we expect that many-body effects
do not strongly perturb the QP-DOS.)
However, there is discrepancy in the BIS part.  A possible assignment is to
identify a shoulder of total QP-DOS seen at $\sim$ \efermi+7eV as the peak
of the BIS at $\sim$ 6.8~eV (as claimed in Ref.~\cite{faleev04}).
Alternatively it is possible that the QP-DOS predicts conduction bands
$\sim 1.5$eV higher than the BIS data.

Fig.~\ref{fig:mno_eps} compares ${\rm{Im}}\epsilon(\omega)$ generated by
\qsgw\ `mode-A' with experiment. The discrepancy looks too large to explain
the difference between the `mode-A' calculation and the
experimental data.  However, if we neglect the difference in absolute
value, we can say that `mode-A' predicts the peak ${\rm
Im}\epsilon(\omega)$ at an energy too high by $\sim$2~eV.  This view is
consistent with the conjecture below for NiO.

\subsubsection{\bf NiO}

The majority-spin QP-DOS of NiO is roughly similar to that of MnO.  However,
the $e_g$ state just below \efermi\ is broadened and carries less weight
(Fig.~\ref{fig:mno_dos}).  That peak is seen in MnO but is lost in this
case, though the deeper $e_g$ peak remains.  The $t_{2g}$ DOS is widened
relative to LDA.  These features are also observed in other beyond-LDA
calculations \cite{massidda97,bengone00}.  The top of valence
consists of O $2p$ states, which hybridize with majority $e_g$ states, and
minority $t_{2g}$ which weakly hybridize.  In the bottom panel, the $d$ DOS
between \efermi\ and \efermi$-$4eV falls in good agreement with XPS; but
the valence DOS width differs from XPS, in contradistinction to MnO.  The
reason can be attributed to the satellite structure contained in the XPS
data: e.g. Takahashi et al.~\cite{takahashib} predict that a satellite
should appear around \efermi$-$9eV.

Turning to the unoccupied states, we can see `mode-A' puts a peak $\sim$1.3
eV too high compared BIS peak at $\sim$\efermi$+$5eV.  On the other hand,
${\rm{Im}}\epsilon(\omega)$ is in rather reasonable agreement with
experiment except for a shift in the first peak by $\sim${2.0}~eV
(`mode-A').  Thus we can distinguish two kinds errors in bandgap: (1.3~eV
in BIS, and 2.0~eV in Im$\epsilon$).  We think both kinds of errors can be
explained by the excitonic effects for $W$ missing in our \qsgw\
calculation.  This is consistent with the \qsgw\ $\epsilon_\infty$ being
underestimated.  Thus our conjecture is: if we properly include the
excitonic contributions to $W$, weights in Im$\epsilon$ will shift to lower
energy, increasing $\epsilon_\infty$.  Self-consistency with such
$W$ should reduce the band gap, simultaneously improving agreement with
BIS and the dielectric function.

\subsection{Fe and Ni}

\begin{figure}[htbp]
\centering
\includegraphics[angle=0,scale=0.4]{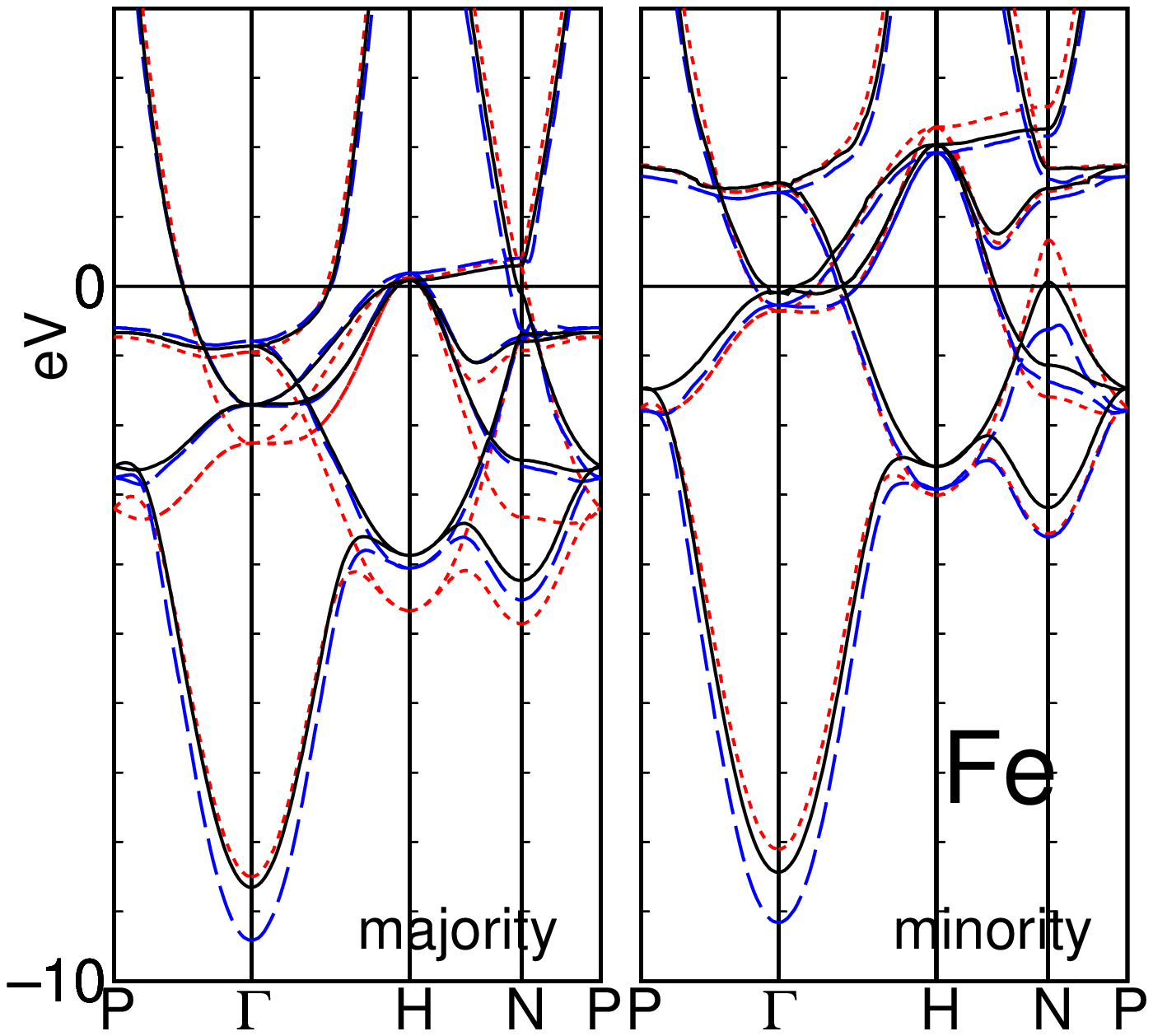}
\includegraphics[angle=0,scale=0.4]{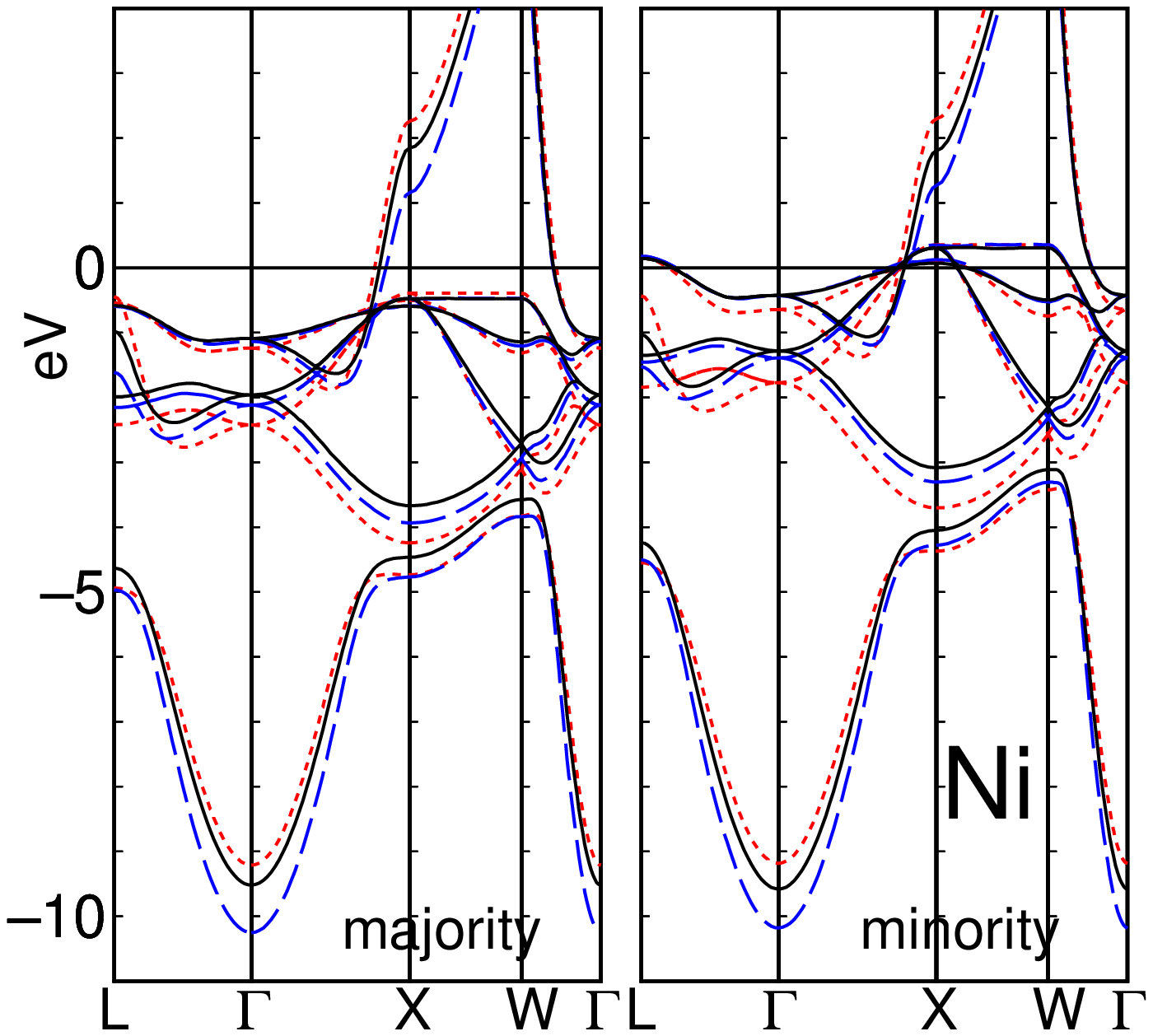}
\caption{(Color online) Energy bands for
Fe and Ni.  
Solid(Black): \qsgw(`mode-A') ; \  
Broken(Blue): `e-only' ; \
Dotted(red): LDA. Calculations
used $12\x12\x12=1728$ ${\bf k}$-points in the 1st BZ.  Comparing to calculations at
other $k$ meshes $8\x8\x8\dots{}14\x14\x14$, we estimate the numerical
convergence is a little better than 0.1 eV.  In this case convergence is
limited by uncertainties in the determination \efermi~\cite{metalnote}.
Inspection of the bands at fine resolution show slight discontinuities at
certain $k$ points.  These occur at times because of difficulties in the
$\Sigma$ interpolation; see Sec.~\ref{sec:siginterpolation} and the smearing
procedure in Sec.~\ref{sec:kint}.  }
\label{fig:fe_band}
\end{figure}

Fig.~\ref{fig:fe_band} shows energy bands for Fe and Ni calculated by LDA,
\qsgw\ `mode-A' and `e-only'.  The two \qsgw\ calculations show similar $d$
band shapes: their widths narrow relative to LDA, moving into closer
agreement with experiment.  On the other hand, the `e-only' calculation
significantly shifts the relative positions of the $s$ and $d$ levels,
depressing the bottom of $s$ band $\sim$1~eV in contradiction to
experiment.  `1shotNZ' results (not shown) are very similar to the `e-only'
case.  This indicates the importance of the charge redistribution due to
the off-diagonal part of \req{eq:veff} in the 3$d$ transition metals.
Yamasaki and Fujiwara~\cite{yamasaki03} presented the `1shot' \gwa\ results
for Fe, Ni, and Cu, Aryasetiawan~\cite{ferdi92} for Ni.  
Both calculations included the 
$Z$ factor ($Z\approx0.8$ for $s$ band, $Z\approx0.6$ for $d$ band) thus
the changes they found relative to LDA are not so large.  Including the $Z$
factor mostly eliminates the unwonted $s$ band shift. However it does so
apparently fortuitously. As `1shot' (and `1shotNZ') should be taken as an
approximation of `e-only', it is wrong to take `1shot' as a better
theoretical prediction than `e-only'.
The calculated spin magnetic moments are listed in Ref.~\cite{mark06qsgw}.
Little difference with experiment is found for Fe (2.2 $\mu_{\rm B}$).
For Ni, \qsgw\ gives 0.7$\mu_{\rm B}$, a little larger than 
the experimental value 0.6 $\mu_{\rm B}$.
This is reasonable because \qsgw\ does not 
include the effect of spin fluctuations. 

\section{summary}

We showed that the \qsgw\ formalism can be derived from a standpoint of
self-consistent perturbation theory, where the unperturbed Hamiltonian is
an optimized noninteracting one.  Then we presented a means to calculate
the total energy based on adiabatic connection, and how it relates to the
density functional framework.

We then showed a number of key points in the implementation of our
all-electron $GW$A and \qsgw.  The mixed basis for $W$ and the
offset-$\Gamma$ method are especially important technical points.  We
presented some convergence checks for the mixed basis, and some results in
various kinds of systems to demonstrate how well \qsgw\ works.  For
insulators, \qsgw\ provides rather satisfactory description of valence
bands; conduction bands are also well described though bandgaps are
systematically overestimated slightly.  We compared the difference of the
\qsgw\ results to eigenvalue-only self-consistent results and to
\oneshotgw\ results in various cases.  

How well eigenvalue-only self-consistency works depends on how correlated
the system is.  In covalent semiconductors LDA eigenfunctions are rather
good, and there is little difference.  The error is no longer small in
Cu$_2$O, Fe, and Ni; finally eigenvalue-only self-consistency fails
qualitatively in NiO.\

We would like to thank F.~Aryasetiawan for giving us his LMTO-ASA $GW$ code,
which formed the basis for development of the present one.
This work was supported by the DARPA SPINS project, and by ONR contract
N00014-02-1-1025.  We thank the Fulton HPC for computational
resources used in this project.

\appendix
\section{How to justify $G^0 W^0$ approximation from $\Sigma[G]$? 
--- Z factor cancellation}
\label{app:zfac}

We explain the ``$Z$ factor cancellation'', 
which is one justification for so-called $G^0 W^0$ approximation.
To the best of our knowledge, it is not clearly discussed in spite of its importance.
Hereafter, $W$ and $\Pi$ denote the dynamical screened Coulomb interaction,
and the proper polarization function without approximation;
$W^0$ and $\Pi^0$ denote the same in RPA.
It gets clearer that the well-balanced treatment between 
the vertex function $\Gamma$ and $G$ to respect
``$Z$ factor cancellation'' is important; thus
the so-called full self-consistent $GW$ \cite{holm98,zein02,weiku02,Biermann03,stan06} 
must be a problematic approximation.
We will use symbolical notations hereafter for simplicity.


As is well known, the exact self energy $\Sigma$ is calculated from
$G$ as $\Sigma = G W \Gamma$ \cite{hedin65}.  $G$ can be written as 
\begin{eqnarray}
G = Z G^0 + \bar{G},
\label{eq:lw}
\end{eqnarray}
where $G^0$ is the QP part of the Green's function, $Z$ is the renomalization
factor, and $\bar{G}$ is the incoherent part.  The incoherent part contains
physically unclear kinds of intermediate states, which 
are not always characterizable by a single-particle propagator.

In (a) below, we consider the $Z$-factor cancellation in 
the calculation of $\Sigma=GW\Gamma$ for given $G$, $W$, and $\Gamma$
and in (b), the $Z$ factor cancellation in $\Pi$.

\setcounter{Alist}{0}
\begin{list}{({\it \roman{Alist}})\,}{\leftmargin 0pt \itemindent 24pt \usecounter{Alist}\addtocounter{Alist}{0}}
\item[(a)]
In the integration of $G W\Gamma$, the
most dominant part is related to the long range static part of $W$, the
$\bfq \to 0$ and $\omega \to 0$ limit of $W(\bfq,\omega)$. 
In this limit, the vertex function becomes
\begin{eqnarray}
\Gamma \to 1- \frac{\partial \Sigma}{\partial \omega} = 1/Z.
\label{eq:wid}
\end{eqnarray}
This is a Ward identity.  Here we need to assume the insulator case.
Then there is a cancellation between $Z$ from $Z G^0$ and $1/Z$ from $\Gamma$.
Under the assumption that $\bar{G}$ is rather structureless,
$\bar{G} W$ may give almost state-independent contributions
(may results in a little changes of chemical potential), 
thus $G W \Gamma \approx G^0 W$ is essentially satisfied.
In the case of metal, there is an additional term 
in right-hand side of \req{eq:wid} due to 
the exsitence of the Fermi surface; then we expect $\Gamma >1/Z$,
see e.g. \cite{nozieres64}; the point (b) below 
should be interpreted in the same manner.
In any cases, we can claim the poorness of fully self-consistent $GW$
which neglect $\Gamma$, as discussed in the follwing.


\item[(b)]
$W$ is given in terms of the proper polarization function $\Pi$
as $W= v(1-v \Pi)^{-1}$. 
$\Pi(1,4) = \bar{\Pi}(1,1;4,4)$, which is written as
\begin{widetext}
\begin{eqnarray}
\bar{\Pi}(1,1';4,4') 
= G_2(1,1';4,4') + G_2(1,1';2,2') I(2,2';3,3') \bar{\Pi}(3,3';4,4') \nonumber \\
= G_2 + G_2 I G_2 + G_2 I G_2 I G_2 + \dots.
\label{eq:barpi}
\end{eqnarray}
\end{widetext}
Here $G_2(1,1'; 2,2') = -i G(1,2) G(2',1')$, and
the duplicated indices are integrated (the Einstein sum rule). 
This looks like an electron-hole ladder diagram, but with the two-particle
irreducible kernel $I(1,1';2,2')$ playing the role of the steps of the ladder.
$G_2$ contains an electron-hole pair excitation 
$\Psi_i(\bfr_1) \Psi_i^*(\bfr_1')\Psi_j^*(\bfr_2) \Psi_j(\bfr_2')
\times (n_j-n_i)\delta(\omega -(\ei-\ej))$
multipled by $Z^2$ in its intermediate state
(imaginary part), because $G_2$ contains $Z G^0 \times Z G^0$.
Here $\ei$ and $\Psi_i$ denote a QPE and QP eigenfunction
included in $G^0$, and $n_j-n_i$ is the occupation number difference.

Let us consider how much the pair excitation  
is included in $\bar{\Pi}$ 
(i.e. how $\bar{\Pi}(1,2;3,4)$ changes when a pair excitation is added or removed)
in its intermediate states.
This means take the derivative of $\bar{\Pi}$
through $G_2$ with respect to $n_j-n_i$ 
(derivative is not through $I$. Such contribution is not two-particle reducible,
that is, not for the intermediate states for $\bar{\Pi}$).

We can show that
\[
\iDelta \Pi(1,4) = \Gamma(1,2,2') \iDelta G_2(2,2';3,3') \Gamma(4,3',3)
\]
;this is derived from \req{eq:barpi} with paying attention to
the matrix notation; $\Gamma(1,2,2')= \frac{1}{1-G_2 I}$ 
and $\Gamma(4,3',3)=\frac{1}{1-I G_2}$ symbolically.
Thus we see that the additional pair excitation (intermediate state)
is included in $\Pi$ with the weight $\frac{1}{Z} \times  Z^2 \times \frac{1}{Z} = 1$
for $\bfq\to{}0, \omega\to{}0$, 
because of \req{eq:wid} (factors ${1}/{Z}$ come from $\Gamma$).
This is the $Z$ factor cancellation mechanism for $\bar{\Pi}$.
Bechstedt et al. demonstrated this in practice \cite{Bechstedt97}
at the lowest level of approximation.

This suggests that $\Pi(1,2) \approx \Pi^0(1,2)=-i G^0 \times G^0$ 
is a reasonable approximation because the derivative 
of $\Pi^0(1,2)$ apparently does not include any $Z$ factor --- thus
the $Z$ factor cancellation is trivially satisfied.
\end{list}


%

In the above discussion, we use the fact that $GW$A is dominated
by the long range part of $W$; we may expect such $Z$ cancellation
somehow occurs even for short-range $W$; but it may be less meaningful.
The above discussion shows why the fully self-consistent $GW$ method is a poor approximation.
Because the vertex function is omitted, $Z$ in $G=ZG^0+\bar{G}$ is not canceled.
$\Sigma= \frac{\delta E_{\rm xc}[G]}{\delta G}$
must be a rigorous formula; however, the series expansion in $G$
should be very inefficient --- it contains rather large cancellations
between terms in the series so as to cancel out the effect of $Z$ as seen in (a) and (b).

On the other hand, the $G^0 W^0$ approximation looks reasonable 
from the viewpoint of (a) and (b), because it 
includes contributions from QPs with correct weights.
From the beginning, this is what we expect from the Landau-Silin 
QP picture.

$Z$ factor cancellation is generally important.
For example, the Bethe-Salpeter equation (BSE) can be described
as the sum of the ladder diagrams; if $G$ is used instead of $G^0$
in the sum, it should give a similarly poor result.

To summarize, it looks more reasonable to calculate $\Sigma[G]$ through 
the QP part $G^0$ contained in $G$. 
That is, $G \to G^0 \to \Sigma$, where we can use $G^0W^0$ 
approximation for $G^0 \to \Sigma$.
From this $\Sigma$, we can calculate a new $G$; this suggest
the self-consistency cycle $G \to G^0 \to G \to G^0 ...$.
The problem is how to extract $G^0$ from $G$; the \qsgw\ method gives
a (nearly) optimal prescription.

Mahan and Sernelius \cite{mahan89} also emphasized the balanced treatment of
the vertex function and $W$.  Their work however, is not directly related to the discussion here.
Their calculation is not based on $GW\Gamma$; instead they use $G^0$, 
and their vertex at $\bfq \to 0$ is not $1/Z$, but unity.
Their formula is based on the derivative of the $G^0$-based total energy 
with respect to the occupation number \cite{quinn58}. 
Their vertex function is identified as the
correction to modify $W$ into 
the effective interaction 
between a test charge and a QP. Their formula
(or originally from Quinn et al. \cite{quinn58})
is related to the discussion in Appendix~\ref{app:etot}.


\section{Can we determine $G$ by total energy minimization?}
\label{app:etot}

The RPA total energy \req{erpac} can be taken as a functional of 
$\veff(\bfr,\bfr')$:  $E^{\rm RPA}$ depends on $\veff(\bfr,\bfr')$ through $G^0$.
Note that the HF part of the total energy does not explicitly include 
QPEs (eigenvalues of $\H0$), but $E^{\rm c,RPA}$ does include them.

In contradiction to the local potential case as, e.g. in the Kohn-Sham
construction of DFT, it is meaningless to minimize 
$E^{\rm RPA}$ with respect to $\veff(\bfr,\bfr')$. 
$\veff(\bfr,\bfr')$ contains degree of freedom 
that can shift QPEs while
keeping the eigenfunctions fixed
(This is realized by adding a potential propotional to
$\psi_i(\bfr) \psi^*_i(\bfr')$). 
Thus it is possible to change only $E^{\rm c, RPA}$ by varying
$\veff(\bfr,\bfr')$ in such a way that QPEs change but not eigenfunctions.
This implies that $E^{\rm c, RPA}$ can be infinite (no lower bound) when
all QPEs are moved to the Fermi energy.



On the other hand, 
it is possible to determine a QPE from 
the functional derivative of $E^{\rm RPA}$
with respect to the occupancy of a state $\Psi_i$.
It gives the QPE as $\ei = \frac{\partial E^{\rm RPA}}{\partial n_i}$
\cite{quinn58}. 
This is in agreement with QPE calculated by $GW$A starting from $\H0$. 
Thus, under the fixed QP eigenfunctions, 
we can determine QPEs self-consistently; we use these $\ei$ in
$E^{\rm RPA}$ and take its derivative 
with respect to $\Psi_i$
to determine the next $\ei$ --- this is repeated until converged. 
This is nothing but the eigenvalue-only self-consistent scheme.
QPE are not determined by the total energy minimization: that is
QPEs are not variational parameters.
Nevertheless it is a self-consistency condition (consistency for the
excitations around the ground state) and it is meaningful.


Therefore, we can calculate $E^{\rm RPA}$ for a given complete 
set of QP eigenfunctions, where QPEs are made self-consistent in
the manner above. It will be possible to minimize this $E^{\rm RPA}$ 
with respect to the set of QP eigenfunctions.
However, such a formalism looks too complicated. 
Further, only the occupied QP eigenfunctions are
included in the Hartree-Fock part of $E^{\rm RPA}$; thus, continuity (smoothness) 
from the occupied eigenfunctions to unoccupied eigenfunctions will be lost.
Thus we think that it is better to choose another possibility, namely to
determine not only QPEs but also the QP eigenfunctions in the
self-consistency cycle, as we do in \qsgw.

\bibliography{qsgwf,lmto,gw}

\begin{thebibliography}{96}
\bibitem[{cor()}]{coretrunc}
\bibinfo{note}{In the $GW$ scheme as we have currently implemented it, CORE
  wave function can be obtained subject to the boundary condition that the
  value and slope vanish at the MT boundary. This procedure do not assure the
  orthogonality between CORE and VAL (the importance of orthogonality is seen
  at $\bfq \to 0$ limit in dielectric function); this can be a source of error
  in the present treatment of CORE.}
\bibitem[{eps()}]{epsnote}
\bibinfo{note}{Both numerator and denominator in the calculation of $\Pi$
  vanish as $\bfk\to{0}$ and $\omega\to{0}$. Thus we need to carefully
  extrapolate to $\bfk\to{0}$ from $\epsilon(\bfk,\omega=0)$. But we found that
  just using $|\bfk| \sim 0.02-0.03$ a.u. is usually good sufficient to
  determine $\epsilon$ to one-tenth of digit for these cases without any
  numerical difficulties.}
\end{thebibliography}

\end{document}